\documentclass[pra,twocolumn,superscriptaddress,floatfix,longbibliography]{revtex4-1}

\newcommand{\ket}[1]{|{#1}\rangle}
\newcommand{\bra}[1]{\langle{#1}|}

\newcommand{\av}[1]{\left\langle#1\right\rangle}
\newcommand{\avs}[1]{\left<#1\right>}

\usepackage{amsfonts}
\usepackage{amsmath}
\usepackage{graphicx}
\usepackage{amssymb}
\usepackage{color}
\usepackage{subfigure}
\usepackage[perpage]{footmisc}
\usepackage{mathtools}
\usepackage{soul}

\begin{document}
\title{Position- and momentum-space two-body correlations in a weakly interacting trapped condensate}

\author{Salvatore Butera}
\affiliation{School of Physics and Astronomy, University of Glasgow, Glasgow G12 8QQ, UK}
\author{David Cl\'ement}
\affiliation{Universit\'{e} Paris Saclay, Institut d'Optique Graduate School,  CNRS, Laboratoire Charles Fabry,  91127 Palaiseau, France}
\affiliation{Institut Universitaire de France (IUF)}
\author{Iacopo Carusotto}
\affiliation{INO-CNR BEC Center and Dipartimento di Fisica, Universit\`a di Trento, I-38123 Povo, Italy}
\begin{abstract}
We investigate the position- and momentum-space two--body correlations in a weakly interacting, harmonically trapped atomic Bose-Einstein condensed gas at low temperatures. The two--body correlations are computed within the Bogoliubov approximation and the peculiarities of the trapped gas are highlighted in contrast to the spatially homogeneous case.
In the position space, we recover the anti--bunching induced by the  repulsive inter--atomic interaction in the condensed fraction localized around the trap center and the bunching in the outer thermal cloud. In the momentum space, bunching signatures appear for either equal or opposite values of the momentum and display peculiar features as a function of the momentum and the temperature. In analogy to the optical Hanbury Brown and Twiss effect, the amplitude of the bunching signal at close-by momenta is fixed by the chaotic nature of the matter field state and its linewidth is shown to be set by  the (inverse of the) finite spatial size of the associated in-trap momentum components. In contrast, the linewidth of the bunching signal at opposite-momenta is only determined by the condensate size. 
\end{abstract}
\maketitle


\section{Introduction\label{Sec:Intro}}

Correlation functions are among the most powerful tools to characterize the properties of light beams in quantum optics and to access the microscopic structure of condensed matter systems. In the former context, the distinction between thermal light (a lamp) vs. coherent light (a laser) and single-photon source (a single two-level emitter) is typically made by looking at the Glauber coherence functions via the statistical properties of suitable photo-detection signals~\cite{loudon2000quantum,QuantumOptics}. 
In the latter context, proximity to a critical point and the onset of an ordered phase are typically encoded in the correlation functions of the order parameter, namely the local magnetization (for the ferromagnetic transition~\cite{Huang}) or the Bose field (for the Bose-Einstein condensation transition~\cite{CCT:CdF,Gunton:PR1968,Barnett_1,Barnett_2,Barnett_3}). 


In spatially large systems, position-space correlations between local observables are typically determined by the bulk properties of the system and are only weakly affected by its unavoidably finite size and by the presence of edges. The situation is completely different for what concerns correlation functions in the reciprocal space, e.g. between different momentum states. The Fourier transform relating position and momentum spaces is in fact a strongly non-local operation and, as such, is strongly affected by the overall size of the system.

The most celebrated example in this sense are the Hanbury Brown and Twiss (HB-T) correlations between different momentum states of an atomic cloud~\cite{schellekens2005, folling2005, ottl2005, perrin2012, dall2013, fang2016, carcy2019}. These studies were inspired by pioneering quantum optics experiments exploiting a subtle relation between the spatial profile of the intensity correlation function of the light detected on Earth and the angular size of a remote star~\cite{HBT1956-sirius}. Along similar lines, the momentum-space correlation function of the atoms showed a non-trivial bunching signal with a momentum-space linewidth determined by the overall size of the cloud. This is to be contrasted with the typical $\delta$-shaped 
form of momentum-space correlations of spatially infinite and homogeneous systems.

While first works on matter wave HB-T physics were restricted to the simplest case of non-interacting atoms of either bosonic~\cite{schellekens2005,ottl2005,dall2013} or fermionic~\cite{jeltes2007comparison} nature, recent developments have started investigating the richer physics of interacting gases. In such systems, the excitation modes have a collective nature and sizable quantum correlations among the elementary constituents are present in the many-body ground state~\cite{bloch2008many,sandrobook}. On one hand, it is natural that signatures of these many-body properties of the bulk should be well visible in the position- and momentum-space correlation functions. On the other hand, we can also expect  that the detailed structure of the momentum-dependence of the correlation function should keep memory of the overall size of the system and, possibly, of its spatial shape.

Motivated by the recent experiment~\cite{Cayla-expHBT-2020}, in this work we report a complete theoretical study of this physics in a simplest model of many-body system that is amenable to {\em ab initio} numerical calculations and analytical insight, but at the same time is rich enough to display a non-trivial physics. We consider a gas a weakly interacting bosonic particles in the presence of a harmonic trapping potential~\cite{sandrobook}. At low temperatures, this system can be described within the Bogoliubov theory based on a macroscopically occupied Bose-Einstein condensate (BEC) and a set of non-interacting bosonic excitation modes, whose nature spans from low-energy collective modes to high-energy single-particles states~\cite{Castin_Houches}. At finite temperature, the thermal population of these excitation modes gives rise to the thermal cloud. Because of interactions, the ground state also contains a sizable quantum depletion, formed by pairs of particles that are excited out of the condensate by virtual two-body collision processes into states with exactly opposite momenta. 

While this picture holds in spatially infinite and homogeneous condensates, several mechanisms were shown to introduce additional features in the momentum space correlation pattern, from quasi-condensation effects in reduced dimensions~\cite{mathey2009}, to different states of bosonic matter in optical lattices~\cite{toth2008}. Here we focus on the consequences of the finite size of a dilute condensate on the linewidth of the different features in the correlation pattern. 
Inspired by the Hanbury Brown and Twiss argument, one of the goals of this work is to assess the relation between the momentum-space linewidth to the inverse physical size of the system as recently explored in~\cite{Cayla-expHBT-2020}.

The structure of the work is the following. In Sec.~\ref{Sec:Syst} we present the physical system under consideration (Subec.~\ref{Sec:Syst}), and we review the basic concepts of the Bogoliubov approximation (Subsecs.~\ref{Sec:Bogo} and \ref{sec:Bogomodes}). These well-known facts are the basis for our study of the spatial shape of the Bogoliubov modes in trapped geometries that is reported in Sec.~\ref{sec:Bogo_trap}. The general theory of two-body correlations within the Bogoliubov approach is reviewed in Sec.~\ref{SubSec:GenTh} and its application to spatially homogeneous systems is summarized in Sec.~\ref{SubSec:Homogeneous}. 

Sec.\ref{Sec:Harmonic} presents the main results of our work. The interplay of quantum antibunching features due to interactions with the bunching features due to the thermal cloud are highlighted in the real-space correlations discussed in Subsec.\ref{SubSec:PosSpace}. The role of the finite system size is even more visible in the momentum-space correlations discussed in Subsec.\ref{SubSec:MomSpace}: a HB-T effect is responsible for strong bunching correlations between neighboring momentum states. The peak value of these correlations is fixed by the Gaussian statistics to the usual HB-T value; the momentum-space linewidth displays interesting behaviours as a function of temperature and momentum, which can be related to the overall size of the condensed and thermal components of the cloud. Similar features are also found in the momentum-space linewidth of the correlation between opposite momentum states. Conclusions are finally drawn in Sec.\ref{Sec:End}.

\begin{figure*}[!t]
\centering
{\subfigure
{\includegraphics[width=0.212\textwidth]{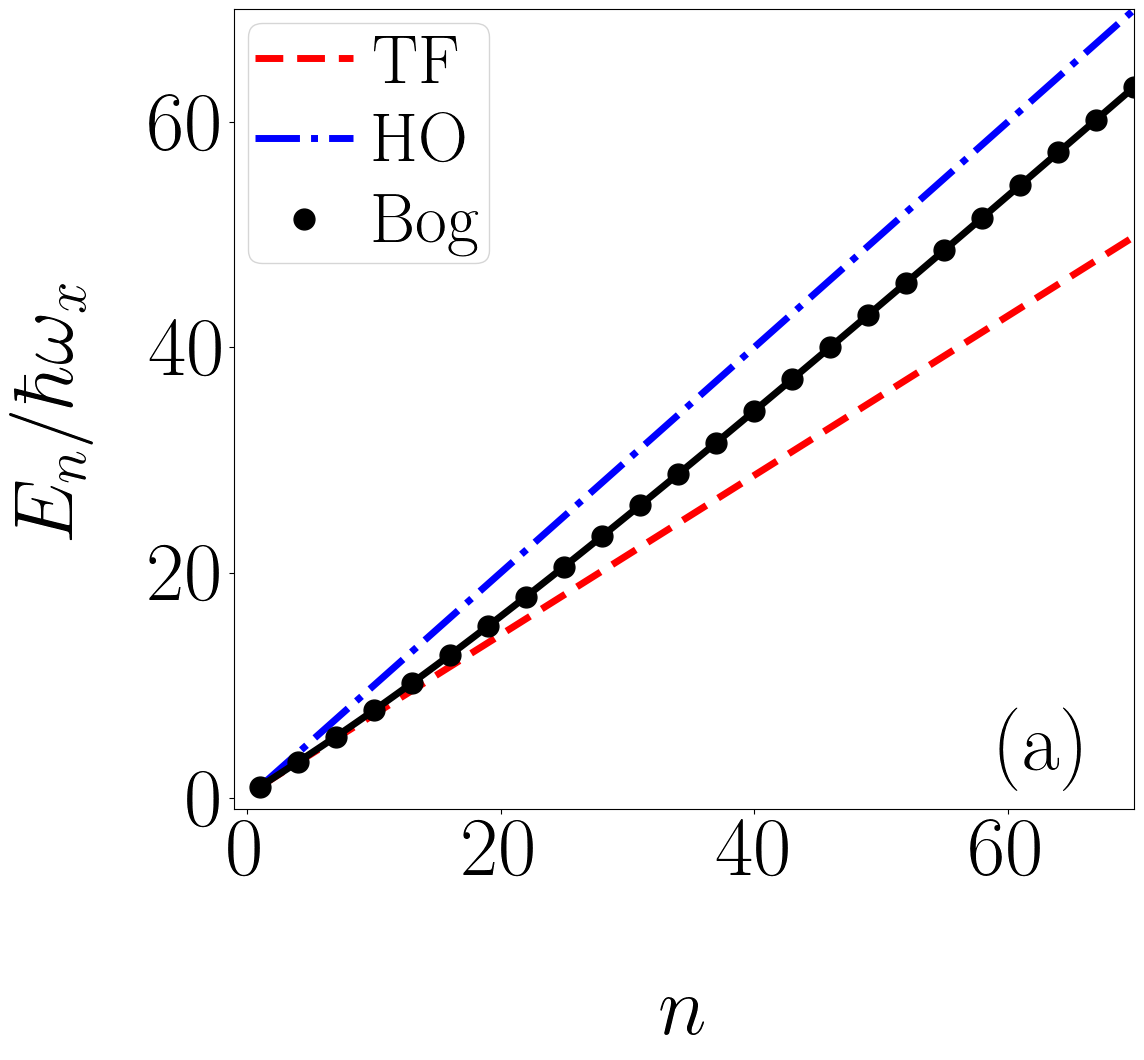}}}
{\subfigure
{\includegraphics[width=0.225\textwidth]{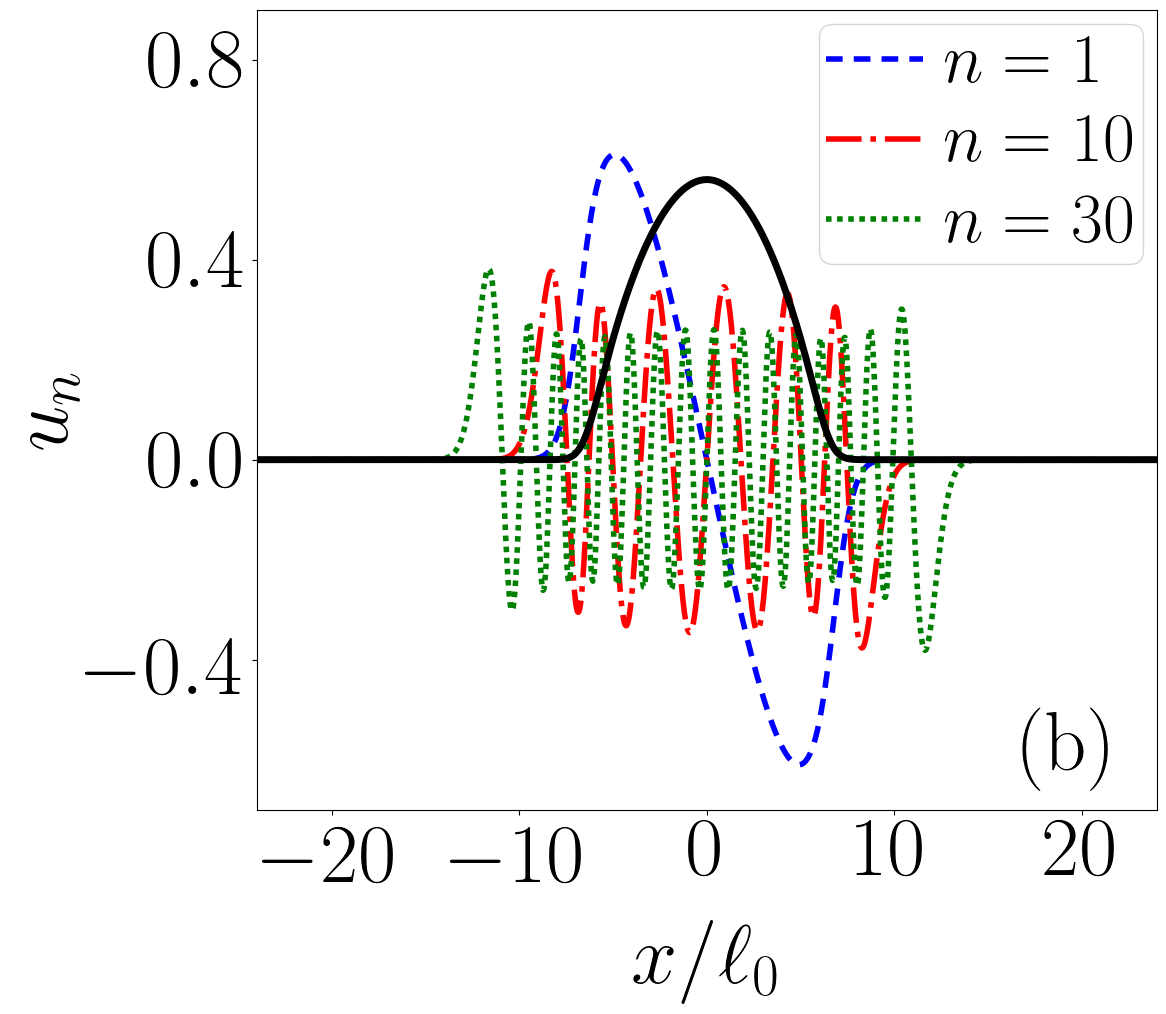}}}
{\subfigure
{\includegraphics[width=0.23\textwidth]{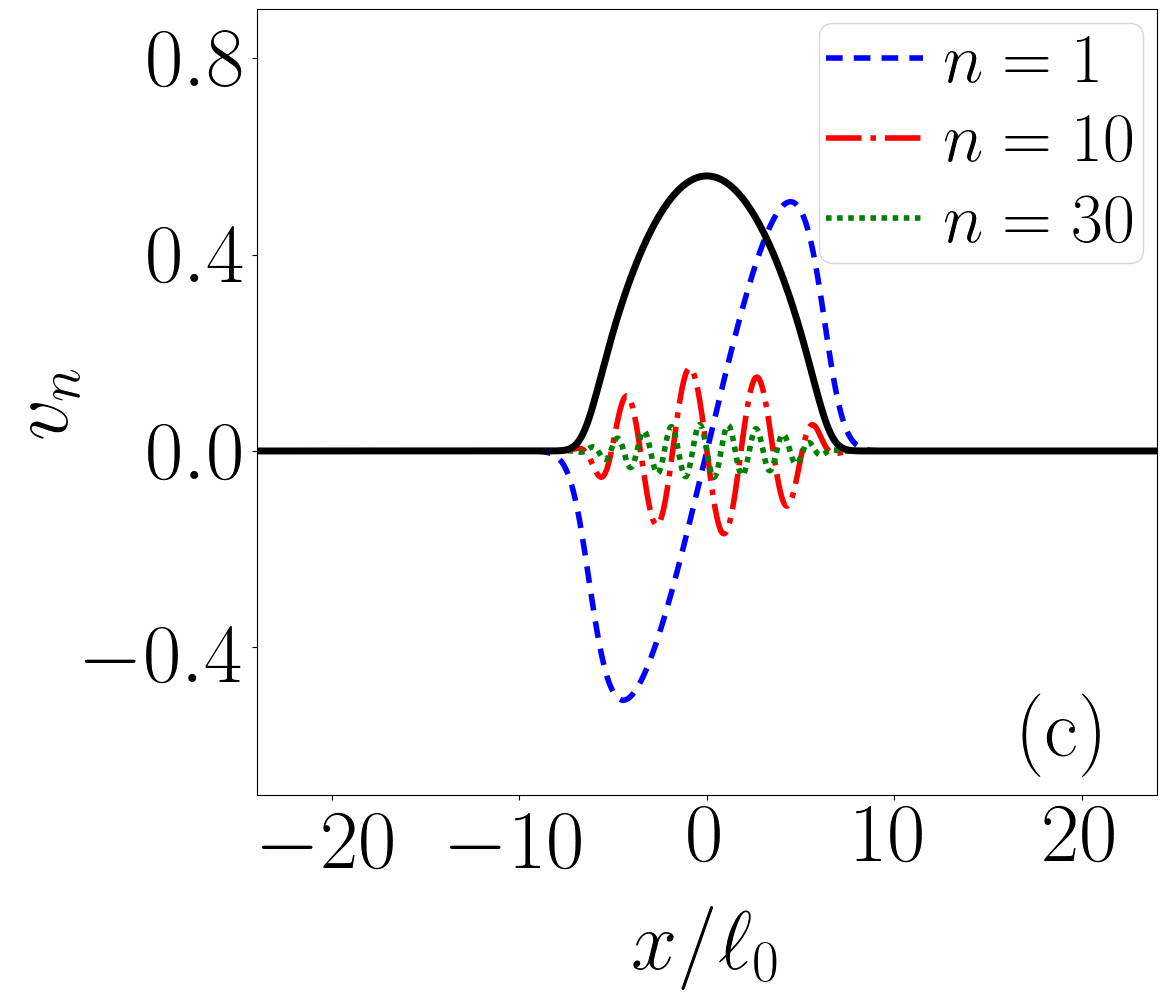}}}
{\subfigure
{\includegraphics[width=0.22\textwidth]{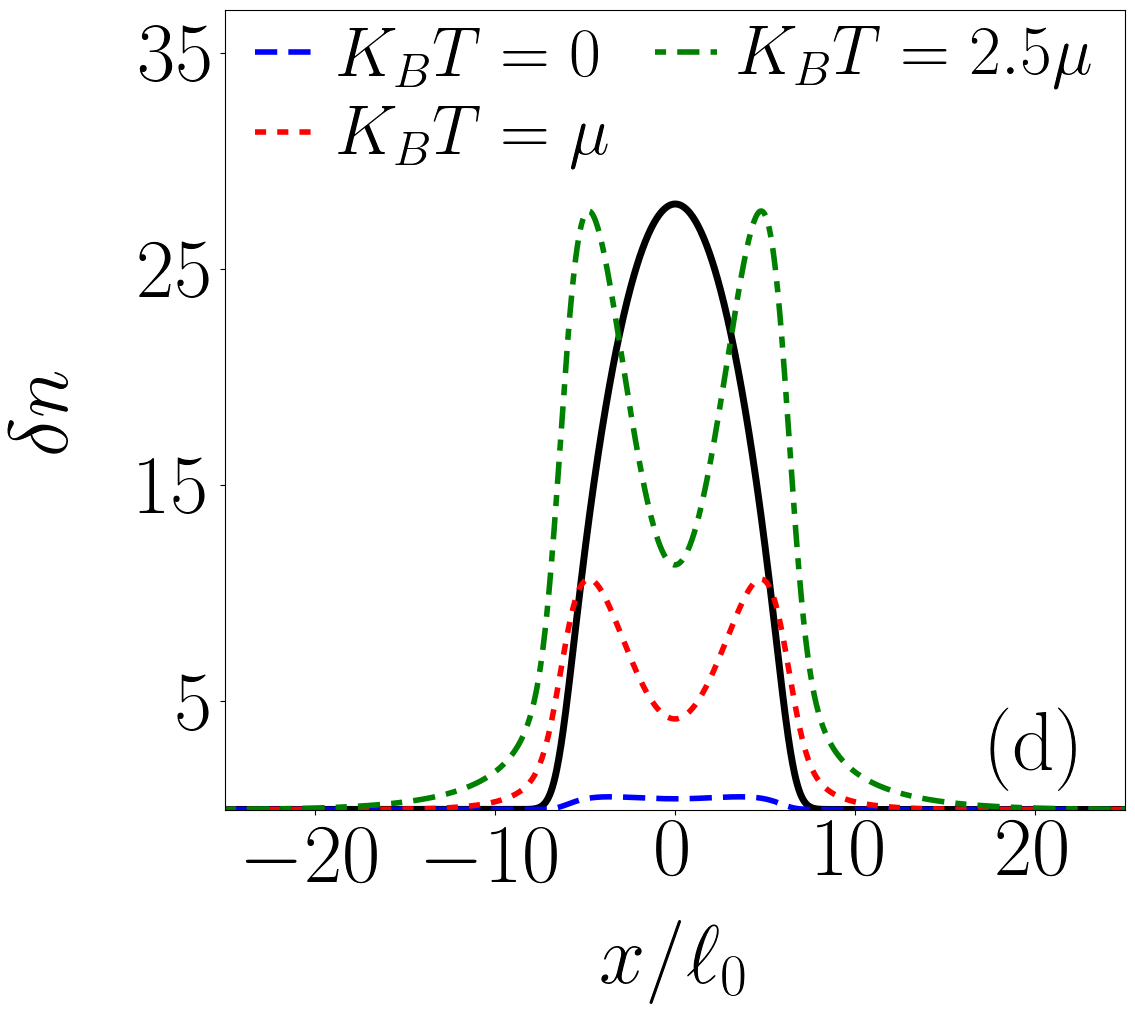}}}

\caption{a) Bogoliubov spectrum in a one-dimensional condensate in the Thomas-fermi regime with $\mu/\hbar \omega_x=11.2\gg 1$. Numerical (black), hydrodynamic Thomas-Fermi limit (dashed red) and harmonic oscillator (dot-dashed blue) predictions for the spectrum of excitation modes as a function of the mode label $n$. b-c) Spatial profiles of the $u_n(x)$ and $v_n(x)$ functions for the $n=1$ (dashed blue line), $n=10$ (dot-dashed red line), $n=30$ (dotted green line) modes of energies $E_{1,10,30}\simeq 1,\,8.59,\,26.0\,\hbar\omega_x $. d) Density of the non-condensed atoms $\delta n$ for three different temperatures, $K_B T=0$ (dashed blue line), $K_B T=\mu$ (dotted red line), $K_B T=2.5\mu$ (dot-dashed green line). The black solid line in panels (b-d) depicts the spatial profile $|\Phi_0(x)|^2$ of the condensate (in units of $\ell_0^{-1}$ in (b-c) and in arbitrary units in (d) for readability sake). The plotted data have been obtained by diagonalizing the Bogoliubov operator in Eg.~\eqref{Eq:BogOperator} in the text, on a grid of $\mathcal{N}_p=2048$ points and an integration box of size $L=80\ell_0$. Here, $\ell_0=\sqrt{\hbar/2m\omega_x}$ is the harmonic oscillator length.}
\label{Fig:1}
\end{figure*}

\section{The physical system and the Bogoliubov description\label{Sec:Syst}}

In this first Section we introduce the physical system under investigation and we review the basic concepts of the Bogoliubov approach that we use in the following. The experienced reader can go straight to Subsec.~\ref{sec:Bogo_trap} where we focus on some intriguing yet less known features of the Bogoliubov modes that are important for the following investigation of the two-body correlations.

\subsection{The physical system}

We consider an ensemble of Bose atoms of mass $m$, pair-wise interacting via an effective contact potential $g\delta(\mathbf{r})$. This describes the low-energy limit of the bare two-body interaction and depends on the scattering length $a_{\rm sc}$ through the coupling strength $g=4\pi\hbar^2a_{\rm sc}/m$. 
The atoms are confined by an external potential $V_{\rm ext}(\mathbf{r})$ which is tightly confined in the transverse $y,z$ plane at a frequency $\omega_\perp$ larger than the chemical potential and the kinetic and interaction energies, so that the system behaves effectively as a quasi-one-dimensional (quasi-1D) gas~\cite{sandrobook}. 

Under this condition, the many-body field operator $\hat{\Psi}(\mathbf{r})$ can be factorized as $\hat{\Psi}(\mathbf{r})=\chi(r_\perp)\hat{\Phi}(x)$, where $\chi(r_\perp)$ is the wavefunction in the radial $r_\perp=(y^2+z^2)^{1/2}$ direction and $\hat{\Phi}(x)$ is the field operator in the longitudinal $x$ direction. 
By integrating over the transverse degrees of freedom we get an effective 1D Hamiltonian in the form
\begin{equation}
\hat{H}=\int{dx\,\hat{\Phi}^\dag(x)\left[\hat{h}+\frac{g_{\rm 1D}}{2}\hat{\Phi}^\dag(x)\hat{\Phi}(x)\right]\hat{\Phi}(x)}.
\label{Eq:1DHamiltonian}
\end{equation}
Here, the single-particle Hamiltonian $\hat{h}= -(\hbar^2/2m) \partial_x^2+V_{\rm ext}(x)$ includes the kinetic energy along $x$ and the confining potential along this direction, $V_{\rm ext}(x)=\frac{1}{2}m\omega_x^2x^2$. The following term describes interparticle interactions whose strength $g_{\rm 1D}=2a_{\rm sc}\hbar\omega_\perp$ is the effective quasi-1D interaction constant for a cylindrically symmetric trap of frequency $\omega_\perp$~\cite{sandrobook}.

\subsection{The Bogoliubov approximation \label{Sec:Bogo}}
Following the number conserving Bogoliubov formalism developed in \cite{Castin-BogNumCons-1998,Gardiner-BogNumCons-1997} and assuming the sample to be short enough along $x$ to avoid {\em quasi}-condensation effects \cite{Petrov_1D}, we split the field operator $\hat{\Phi}(x,t)$ into the \emph{condensed} component describing atoms occupying the single particle condensate state and the \emph{non-condensed} component that accounts for the population of the excited single-particle states,
\begin{equation}
	\hat{\Phi}(x)=\Phi_{0}(x) \hat{a}_{0}+\delta\hat{\Phi}(x).
	\label{Eq:BogAns}
\end{equation}
Here the operator $\hat{a}_{0}$ annihilates a particle from the condensate mode $\Phi_0(x)$, while $\delta\hat{\Phi}(x)$ annihilates a non-condensed particle at position $x$. The process by which a particle is transferred from the non-condensed into the condensate component is described by the action of the operator $\hat{\Lambda}(x)$, defined as
\begin{equation}
	\hat{\Lambda}(x)=\frac{1}{\sqrt{N} }\hat{a}_{0}^\dag\,\delta\hat{\Phi}(x).
\label{Eq:Lambda}
\end{equation}
Here $N$ is the total number of particles, that differs from the (average) number of condensate atoms $N_{0}\equiv\avs{\hat{a}_{0}^\dag\hat{a}_{0}}$ by the amount $\delta N\equiv\int{dx\, \av{\hat{\Phi}^\dag(x)\hat{\Phi}(x)}}$. In the weakly-interacting limit where $N\to \infty$ and $g_{1D}\to 0$ at a fixed mean-field energy $N g_{1D}$, one has $\delta N\sim 1$ and one can adopt a perturbative approach in $\delta N/N$. Note that in this limit the critical temperature for condensation is pushed to high temperatures (strictly speaking to infinity), while the chemical potential $\mu$ remains constant as it is related to the product $N g_{1D}$. 

The Gross-Pitaevskii theory accounts for the zeroth order term of this expansion, while the Bogoliubov theory of non-interacting excitations living on top of the condensate accounts for the first order correction. 
At the lowest non-trivial level in this perturbative expansion, the time evolution of $\hat{\Lambda}(x,t)$ and of its hermitian conjugate $\hat{\Lambda}^\dag(x,t)$ is governed by the Bogoliubov-de Gennes equations \cite{Castin-BogNumCons-1998}
\begin{equation}
	i\hbar\frac{d}{dt}
	\begin{pmatrix}
    \hat{\Lambda}\\
	\hat{\Lambda}^\dag
  \end{pmatrix}=\mathcal{L}
  	\begin{pmatrix}
    \hat{\Lambda}\\
	\hat{\Lambda}^\dag
  \end{pmatrix}= \begin{pmatrix}
    L_{QQ}     & L_{QQ^*}\\
    L_{Q^*Q}      & -L_{QQ}
  \end{pmatrix} \begin{pmatrix}
    \hat{\Lambda}\\
	\hat{\Lambda}^\dag
  \end{pmatrix} ,
  \label{Eq:BogOperator}
\end{equation}
where the (operator-valued) components of the Bogoliubov operator $\mathcal{L}$
are defined as
\begin{subequations}
\begin{align}
	L_{QQ}&=\left[ H_{\rm GP}+N g_{1D}\, Q|\Phi_0(x,t)|^2 Q-\mu\right],\label{Eq:LQQ}\\
	L_{QQ^*}&=N g_{1D}\, Q\Phi_0^2(x,t) Q^*,\label{Eq:LQQ*}\\
	L_{Q^*Q}&=\left(L_{QQ^*}\right)^*.\label{Eq:LQ*Q}
\end{align}
\label{Eq:LComponents}
\end{subequations}
At this level of approximation, in Eq.~\eqref{Eq:LQQ}, $\Phi_0(x)$ coincides with the (normalized) ground state solution of the Gross-Pitaevskii equation \cite{sandrobook} and $H_{\rm GP}=-\hbar^2\partial_x^2/2m+N g_{1D} |\Phi_0(x,t)|^2+V_{\rm ext}(x)$ is the Gross-Pitaevskii Hamiltonian. We denote the Thomas-Fermi radius of the Gross-Pitaevskii ground-state by $L_{\rm bec}$ in the following. In Eqs.~\eqref{Eq:LComponents}(a-c), the operator $Q\equiv \mathbb{I}-\ket{\Phi_0}\bra{\Phi_0}$ is the projector onto the non-condensed component, that is onto the Hilbert sub-space spanned by all single particle excited states. 

\subsection{The Bogoliubov Hamiltonian and its ground-state}
\label{sec:Bogomodes}

The $\hat{\Lambda}(x)$ operator can be conveniently expressed in the basis of the Hilbert space composed by the positive norm and positive frequency eigenvectors $[u_n(x),v_n(x)]^T$ (with $n\in \mathbb{N}$) of $\mathcal{L}$ and the associated negative norm and negative frequency ones $[v_n^*(x),u_n^*(x)]^T$~\cite{Castin-BogNumCons-1998,Castin_Houches}.
In terms of this basis, we can expand the non-condensed operators as
\begin{equation}
	\hat{\Lambda}(x)=\sum_{n\in (+)}\left(\hat{b}_n u_n(x)+\hat{b}_n^\dag v_n^*(x)\right),
\label{Eq:LambdaExpansion}
\end{equation}
where the sum runs over the positive norm modes only. Note that for a trapped condensate, all $u_n(x)$ and $v_n(x)$ can be taken as real.

By combining this expansion with the decomposition  Eq.~\eqref{Eq:BogAns} of the field operator, the Hamiltonian of the system in Eq.~\eqref{Eq:1DHamiltonian} can be conveniently written, to order $\mathcal{O}(1)$ in the particle number $N$, in the diagonal form
\begin{equation}
	\hat{H}=E_0(N)+\sum_n{E_n\hat{b}_n^\dag\hat{b}_n},
\label{Eq:HamiltonianDiag}
\end{equation}
where $E_0(N)$ is the ground state energy (see Eq.~(71) in \cite{Castin-BogNumCons-1998} for the explicit form). The $\hat{b}_n$ and $\hat{b}^\dag_n$ operators satisfy bosonic commutation rules, $[\hat{b}_m,\hat{b}^\dagger_n]=\delta_{m,n}$ and physically correspond to the destruction and creation of quanta of excitation in the different excitation modes of the condensate. The Bogoliubov ground state $\ket{0}_{\rm bog}$ is the (unique) state that is annihilated by all quasi-particle operators $\hat{b}_n$,
\begin{equation}
\hat{b}_n\ket{0}_{\rm bog}=0 \qquad \forall n. 
\end{equation}
The Bogoliubov ground state $\ket{0}_{\rm bog}$ is different from $\ket{\Phi_0}$ in that it contains an admixure of (non-condensed) single-particle excited states, which are referred to as the \emph{quantum depletion}. Note that the Bogoliubov theory do not include terms beyond quadratic order in the Hamiltonian (\ref{Eq:HamiltonianDiag}), which physically means that the different excitation modes are assumed to be not interacting.

%

\subsection{The Bogoliubov modes of a trapped condensate}
\label{sec:Bogo_trap}

In the Bogoliubov approximation, the structure of the excitations of a trapped condensate are readily obtained by diagonalizing the Bogoliubov operator \eqref{Eq:BogOperator}. We have numerically diagonalized the Bogoliubov operator on a one-dimensional lattice of $\mathcal{N}_p$ points uniformly spaced by $dx=L/\mathcal{N}_p$. The kinetic energy is implemented by evaluating the spatial second derivative via a fourth order finite difference scheme \cite{Fornberg}. Care has been paid to ensure convergence of the results on the IR side against the integration box size $L$ and on the UV side against the lattice spacing $dx$. This diagonalization provides us with both the eigen-energies, i.e. the spectrum, and the eigenmodes, i.e. the Bogoliubov excitations modes, that we now briefly describe.

In Fig.~\ref{Fig:1}(a) we report the numerical result for the spectrum of the Bogoliubov operator $\mathcal{L}$ for a harmonically trapped condensate of chemical potential ${\mu=11.2~\hbar \omega_x}$. We compare it with the analytical solutions obtained in the Thomas-Fermi (red dashed) and non-interacting (blue dot-dashed) limits. The former reproduces well the low-energy modes of {\em collective} --phononic-- nature. The latter recovers the high-energy part of the spectrum, where modes have a {\em single particle} nature~\cite{sandrobook,Castin_Houches}. The crossover between the two regimes is set by the interaction energy $\mu \simeq g_{1D} n(0)$, where $n(x)\equiv N|\Phi_0(x)|^2$ is the condensate density. 

In Figs.~\ref{Fig:1}(b,c) we show the numerical results for a few of the functions $u_n(x)$ and $v_n(x)$ associated with the two types of Bogoliubov excitation modes. We have plotted the solutions for the $n=1,\,10,\,30$ modes of energy $E_{1,10,30}\simeq 1,\,8.59,\,26.0\,\hbar\omega_x $. These values were chosen to highlight the structure of the Bogoliubov modes in the {\em collective} ($n=1$), intermediate ($n=10$) and {\em single-particle} ($n=30$) part of the energy spectrum. At low energies, i.e., small values of $n$, where the excitations have a {\em collective} nature, the $u_n,v_n$ have similar amplitudes. In the opposite regime of {\em single-particle} excitations, i.e., at large values of $n$, the amplitude of the $v_n$ tends to vanish while that of the $u_n$ remains almost unaffected. 

Also the spatial profiles of both the $u_n$ and $v_n$ functions change drastically with $n$. These modifications are central to understand the structure of the two-body correlations that we are going to discuss in the next Sections. More specifically, the $u_n(x)$ and $v_n(x)$ functions of low-energy modes (e.g. the blue dashed line for the $n=1$ dipole mode in Figs.~\ref{Fig:1}(b,c)) have a similar shape and are localized within the condensate (whose density profile is indicated by the black solid line). This is because the {\em collective} character of low-energy modes requires the presence of the underlying condensate. In the opposite regime of large energies and {\em single-particle} excitations, the spatial extensions of the $u_n(x)$ and $v_n(x)$ strongly differ from one another. On the one hand, the functions $u_n(x)$ of high-energy modes (e.g. the green dotted line for $n=30$ in Fig.~\ref{Fig:1}(b)) display a standing wave profile with a relatively uniform envelope which extends well beyond the density profile of the condensate. This is because highly energetic single particles can freely climb along the sides of the harmonic trap outside the condensate. On the other hand, the $v_n(x)$ functions (red dash-dotted and green dotted lines in Fig.~\ref{Fig:1}(c)) have a significantly non-zero value only in the condensate region. The envelope of their standing wave profile smoothly reaches the condensate edge but is largest around the trap center: This feature can be ascribed to the non-homogeneous density profile of the condensate and to the way the collective or single-particle character of a mode is related to the local density: in a trap, a given excitation mode of frequency $\omega$ has a more collective nature, and thus a larger $v_n(x)$ in the central high-density region where the local interaction energy $g_{1D} n(x)$ is larger.  

As a final remark, we show in Fig.~\ref{Fig:1}(d) the spatial profile of the non-condensed fraction given, to order $O(1)$, by 
\begin{equation}
\delta n(x) = \av{\hat{\Lambda}^\dag(x)\hat{\Lambda}(x)}.
\end{equation}
First of all, the total number of non-condensed particles increases with the temperature, as expected because the temperature promotes particles outside the ground-state. While the quantum depletion at $T=0$ is relatively flat up to the edge of the condensate, the thermal component visible at higher $T$'s extends well beyond the condensate and is suppressed at the center of the trap by the repulsive effect exerted by the condensate (see the extra term $\sim|\Phi_0(x)|^2$ in Eq.~\eqref{Eq:LQQ}).

\section{Two-body correlations: general theory\label{SubSec:GenTh}}

After having reviewed the basic concepts of the Bogoliubov theory and presented the spatial structure of the excitation modes in a harmonically trapped geometry, we are now in a position to attack the core problem of this work, namely the two-body correlations in both the position- and momentum-space. In this Section, we outline the approach to calculate the correlation functions in the Bogoliubov approximation.

To maintain full generality at this stage, we consider a generic two-body correlation function of the form
\begin{equation}
	G^{(2)}(s_1,s_2)=\av{\hat{\Phi}^\dag(s_1)\hat{\Phi}^\dag(s_2)\hat{\Phi}(s_2)\hat{\Phi}(s_1)},
\label{Eq:G2gen_def}
\end{equation}
where $s_1$ and $s_2$ are generic variables either in the position- or the momentum-space.

Up to order $\mathcal{O}(1)$ in the particle number $N$, Eq.~\eqref{Eq:G2gen_def} can be expanded as
\begin{multline}
	G^{(2)}(s_1,s_2)=N^2\, G^{(2)}_{{\Phi_0}{\Phi_0}}(s_1,s_2)+N\, G^{(2)}_{{\Phi_0}\Lambda}(s_1,s_2)\\
		+G^{(2)}_{\Lambda\Lambda}(s_1,s_2),
\label{Eq:G2gen_exp}
\end{multline}
where the terms proportional to $G^{(2)}_{{\Phi_0}{\Phi_0}}(s_1,s_2)$, $G^{(2)}_{{\Phi_0}\Lambda}(s_1,s_2)$ and $G^{(2)}_{\Lambda\Lambda}(s_1,s_2)$ are of different orders in $N$ and account for different type of correlations. The first term $G^{(2)}_{{\Phi_0}{\Phi_0}}$ describes the trivial correlations between particles in the condensate. Fluctuations on top of the condensate are captured by the following terms: The second term proportional to $G^{(2)}_{{\Phi_0}\Lambda}$ describes correlations between particles in and out of the condensate. The third term proportional to $G^{(2)}_{\Lambda\Lambda}$ describes correlations between the non-condensed particles. At the level of the Bogoliubov theory, they take the explicit form
\begin{widetext}
\begin{eqnarray}
	G^{(2)}_{{\Phi_0}{\Phi_0}}(s_1,s_2)&=&\left|{\Phi_0}(s_1)\right|^2\,\left|{\Phi_0}(s_2)\right|^2,\label{Eq:Ggen_pp}\\
	G^{(2)}_{{\Phi_0}\Lambda}(s_1,s_2)&=&\left[-\left|{\Phi_0}(s_1)\right|^2\left|{\Phi_0}(s_2)\right|^2+ \left(\left|{\Phi_0}(s_1)\right|^2\av{\hat{\Lambda}^\dag(s_2)\hat{\Lambda}(s_2)}+\left|{\Phi_0}(s_2)\right|^2\delta\av{\hat{\Lambda}^\dag(s_1)\hat{\Lambda}(s_1)}\right)\right. \nonumber\\ 
	&+&\left.\left({\Phi_0^*}(s_1){\Phi_0^*}(s_2)\avs{\hat{\Lambda}(s_2)\hat{\Lambda}(s_1)}+ {\Phi_0^*}(s_1){\Phi_0}(s_2)\avs{\hat{\Lambda}^\dag(s_2)\hat{\Lambda}(s_1)}+c.c.\right)\right],\label{Eq:Ggen_pL}\\
	G^{(2)}_{\Lambda\Lambda}(s_1,s_2)&=&\avs{\hat{\Lambda}^\dag(s_1)\hat{\Lambda}^\dag(s_2)\hat{\Lambda}(s_2)\hat{\Lambda}(s_1)},
	\label{Eq:Ggen_LL}
\end{eqnarray}
\end{widetext}
where {\it c.c.} indicates the complex conjugation operation. 

In the position space, the fact that the  different components are not spatially separated makes correlations to be dominated by the $G_{{\Phi_0}\Lambda}^{(2)}(x_1,x_2)$ term in \eqref{Eq:Ggen_pL}. For this reason we will focus on this component when considering the position space correlations. Note that this term is proportional to the correlation function of the density fluctuations, namely the connected component of the density-density correlation function 
\begin{equation}
    {G}^{(2)}_{\rm c}(x_1,x_2)=\langle : \hat{n}(x_1) \hat{n}(x_2) : \rangle - \langle \hat{n}(x_1) \rangle \langle \hat{n}(x_2) \rangle\,.
\end{equation}

In the momentum space, instead, there is a clear separation of scales between the condensate that lives in low-momentum states up to $k\simeq 1/L_{\rm bec}$ and the non-condensed fraction that extends up to much higher momenta determined by the temperature or the inverse healing length of the condensate. This separation allows to separately identify the three terms in Eqs.~(\ref{Eq:Ggen_pp}-\ref{Eq:Ggen_LL}).
The correlations that are of interest for the present work involve modes at $k_1,k_2$ located outside the condensate and are described by the highest-order $G_{\Lambda\Lambda}^{(2)}(k_1,k_2)$ term. In the following of this work we will focus on this term for the momentum-space correlations.

Given the quadratic form of the Bogoliubov Hamiltonian, the thermal equilibrium state has a Gaussian form at any temperature and the Wick expansion is exact. As a consequence, the quartic correlator $G^{(2)}_{\Lambda\Lambda}(s_1,s_2)$ in Eq.\eqref{Eq:Ggen_LL} can be expanded in terms of products of second order correlators as:
\begin{equation}
	G^{(2)}_{\Lambda\Lambda}(s_1,s_2)=G^{(2)}_{\Lambda\Lambda,{\rm N}}(s_1,s_2)+G^{(2)}_{\Lambda\Lambda,{\rm A}}(s_1,s_2),
\label{Eq:Wick}
\end{equation}
where
\begin{eqnarray}
G^{(2)}_{\Lambda\Lambda,{\rm N}}(s_1,s_2)&=&\left|G^{(1)}(s_1,s_2)\right|^2+ \nonumber \\ &+&G^{(1)}(s_1,s_1)G^{(1)}(s_2,s_2),\label{Eq:G2N}\\
G^{(2)}_{\Lambda\Lambda,{\rm A}}(s_1,s_2)&=&\left|A^{(1)}(s_1,s_2)\right|^2,\label{Eq:G2A}
\end{eqnarray}
involve products of respectively the \emph{normal} and \emph{anomalous} averages of the Bogoliubov operator:
\begin{eqnarray}
 G^{(1)}(s_1,s_2)&=&\av{\hat{{\Lambda}}^\dag(s_1)\hat{{\Lambda}}(s_2)} \\ A^{(1)}(s_1,s_2)&=&\av{\hat{{\Lambda}}(s_1)\hat{{\Lambda}}(s_2)}\,.
\end{eqnarray}

In position space, we can make use of the expansion in Eq.~\eqref{Eq:LambdaExpansion} to write these expectation values in terms of the Bogoliubov modes as
\begin{eqnarray}
 G^{(1)}(x_1,x_2)&=&\sum_n \left[(1+N_n)v_n(x_1) v_n(x_2)+\right. \nonumber \\&+&\left.{N_n u_n(x_1) u_n(x_2)}\right], \label{eq:N(1)} \\
 A^{(1)}(x_1,x_2)&=&\sum_n \left[(1+N_n)u_n(x_1) v_n(x_2) +\right. \nonumber \\&+&\left.{N_n v_n(x_1) u_n(x_2)}\right], \label{eq:A(1)} 
\end{eqnarray}
where
\begin{equation}
N_n = \frac{1}{e^{\beta \hbar \omega_n}-1} 
\end{equation}
is the thermal occupation of the excitation mode of frequency $\omega_n$. 

Analogous expressions can be straightforwardly written in the momentum-space by defining the Fourier transforms of the Bogoliubov mode functions $u_n(k)$ and $v_n(k)$
\begin{equation}
    [u_n,v_n](k) = \frac{1}{\sqrt{L}} \int\!dx\,[u_n,v_n](x)\,e^{-ikx}.
\end{equation}
where $L$ is the size of the integration box. These write
\begin{eqnarray}
 	G^{(1)}(k_1,k_2)&=&\sum_n\left[\left(1+N_n\right)v_n^*(k_1)v_n(k_2)+\right.\\
		&+&\left.{N_n u_n^*(k_1)u_n(k_2)}\right],
\label{Eq:LdL_q} \\
	A^{(1)}(k_1,k_2)&=&
		\sum_n\left[{\left(1+N_n\right)}u_n(k_1)v_n(k_2)+\right.\\
		&+&\left.{N_n v_n(k_1)u_n(k_2)}\right].
\label{Eq:LL_q}
\end{eqnarray}
and can be directly used to evaluate $G^{(2)}_{\Lambda\Lambda}(k_1,k_2)$ using \eqref{Eq:Wick}.

\begin{figure*}[!t]
\centering
{\subfigure
{\includegraphics[width=0.3\textwidth]{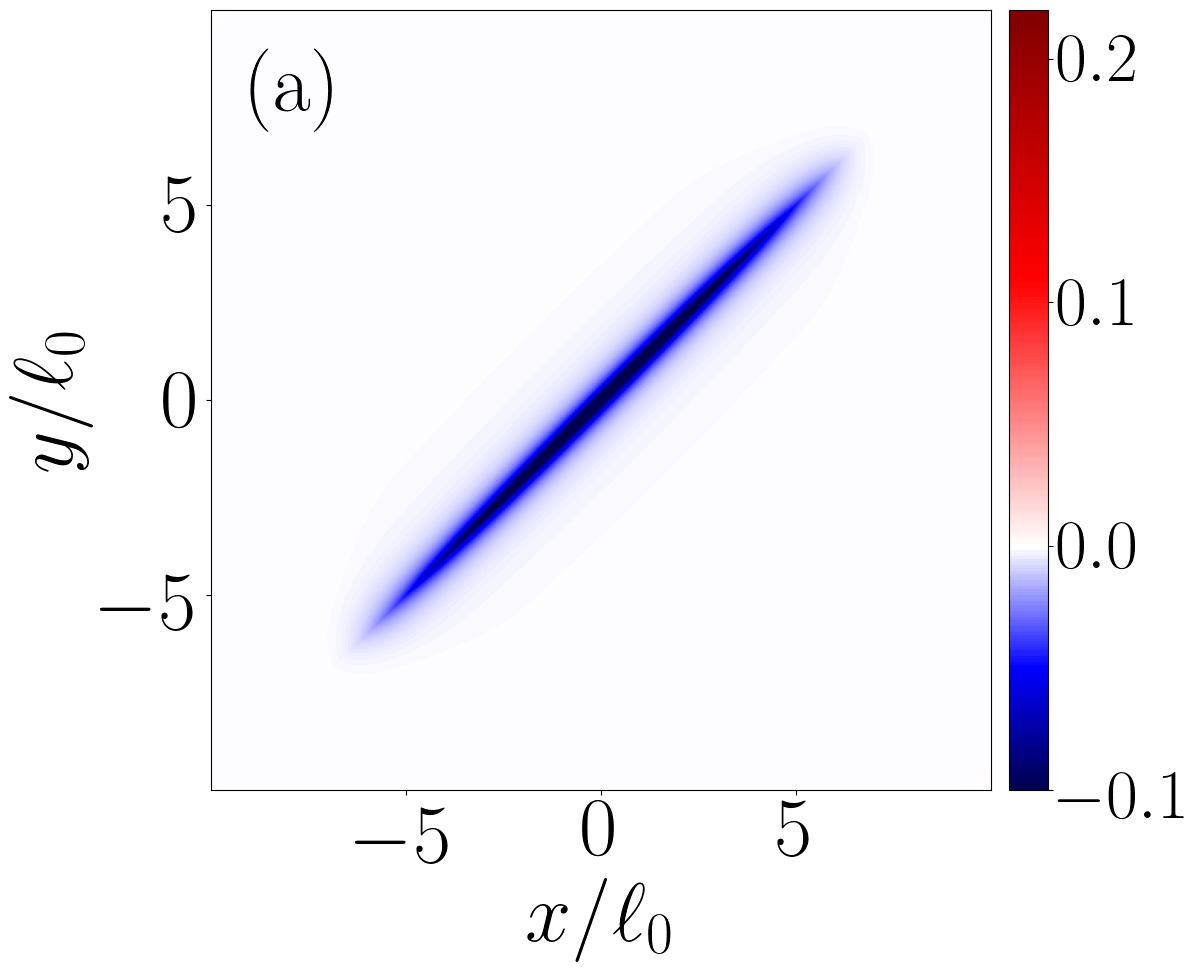}}}\qquad
{\subfigure
{\includegraphics[width=0.3\textwidth]{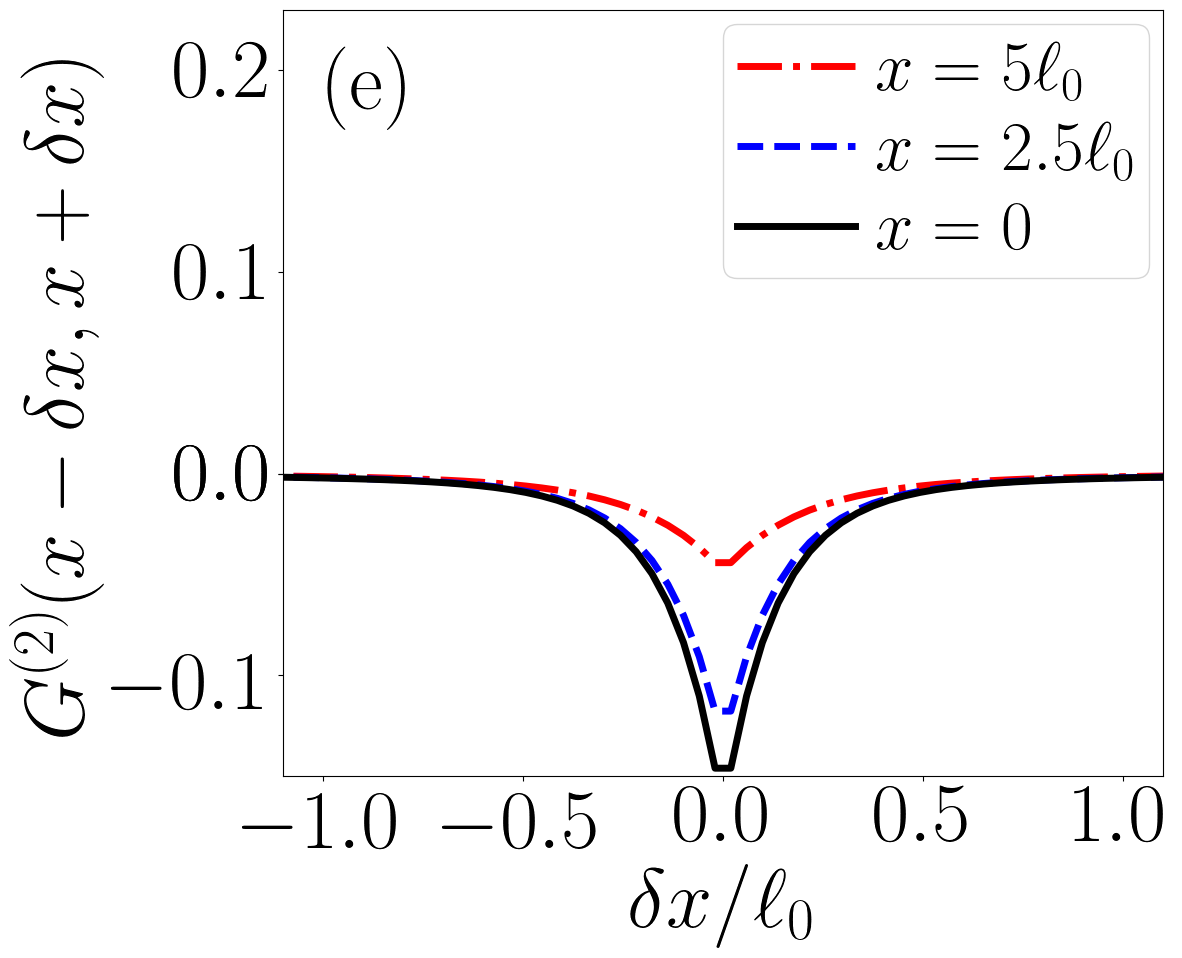}}}\\
{\subfigure
{\includegraphics[width=0.3\textwidth]{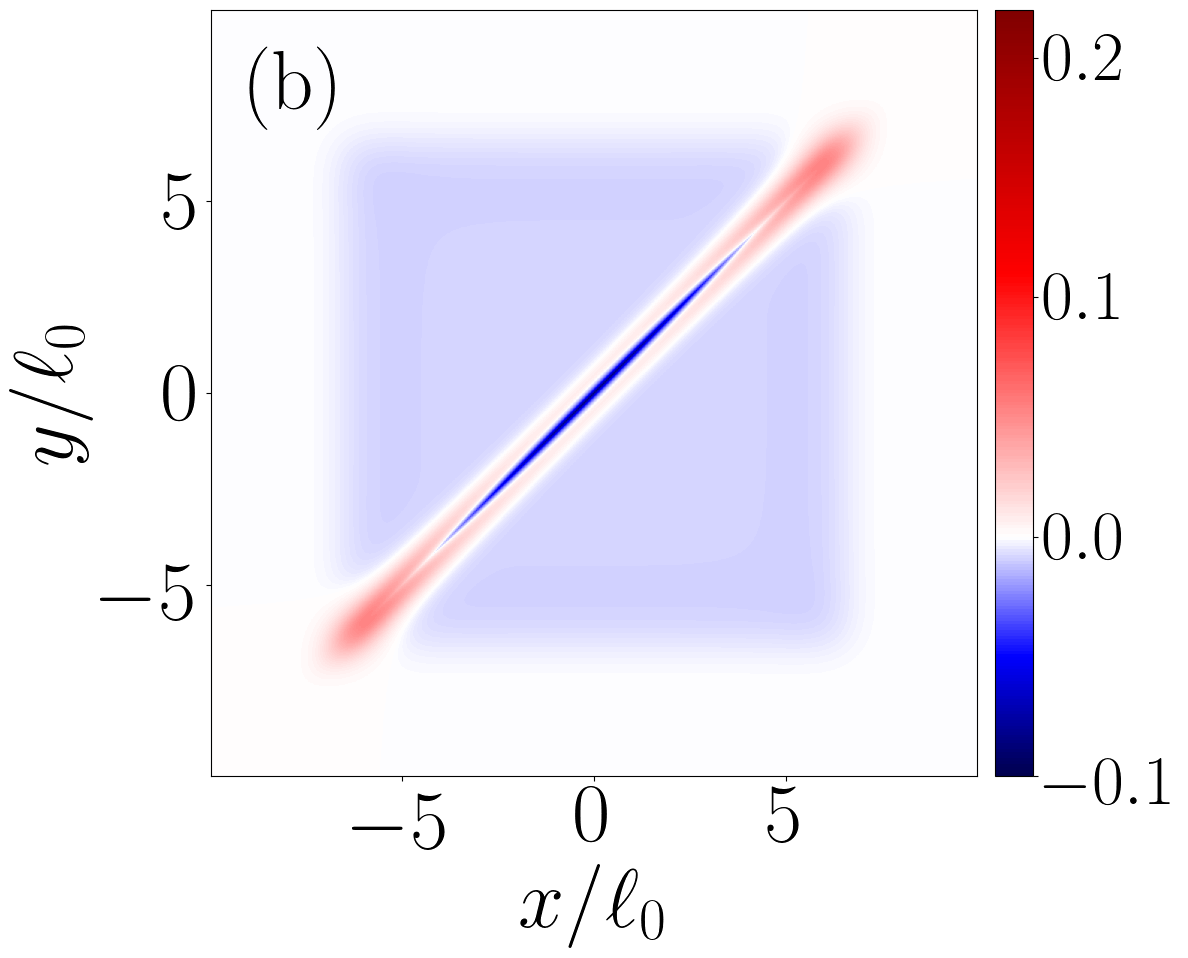}}}\qquad
{\subfigure
{\includegraphics[width=0.3\textwidth]{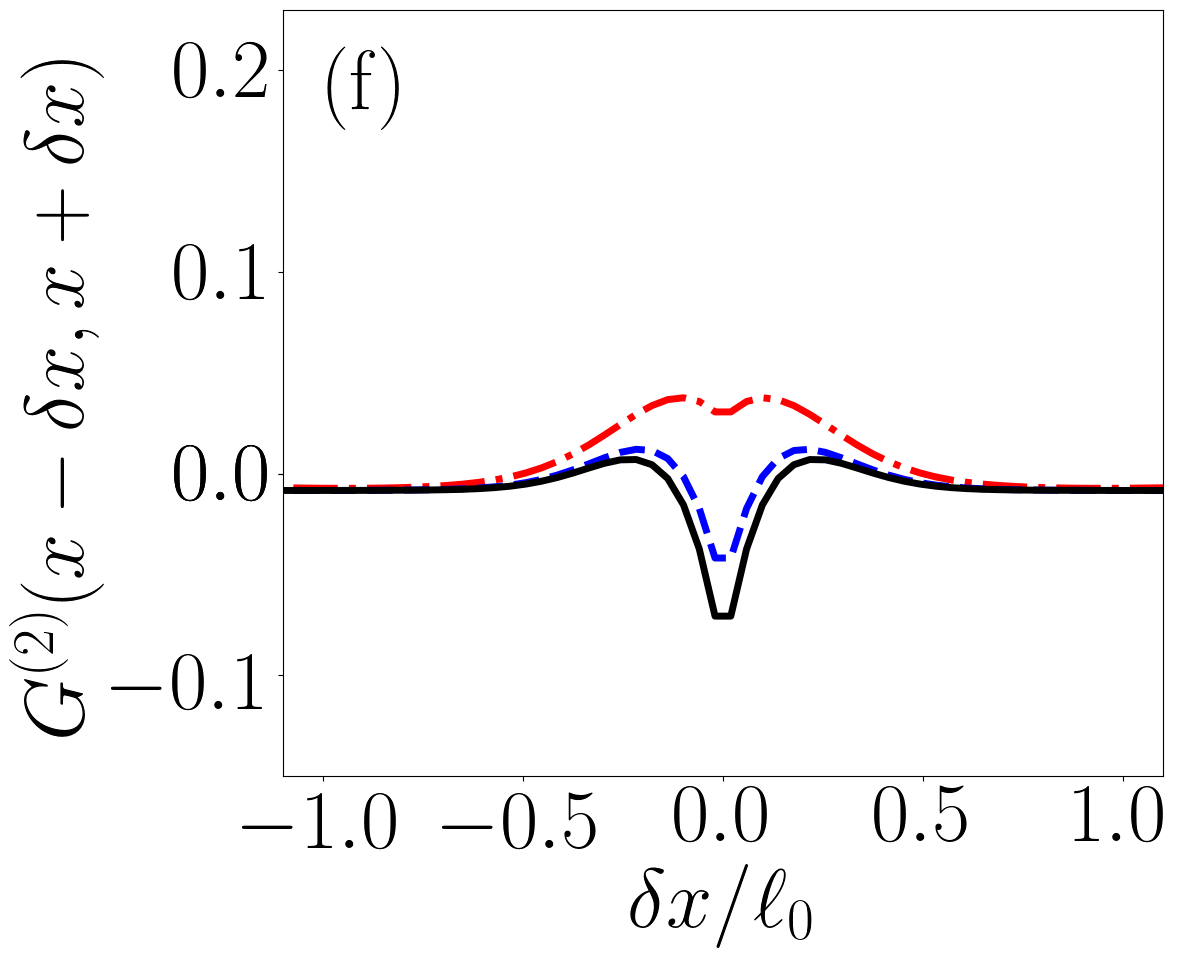}}}\\
{\subfigure
{\includegraphics[width=0.3\textwidth]{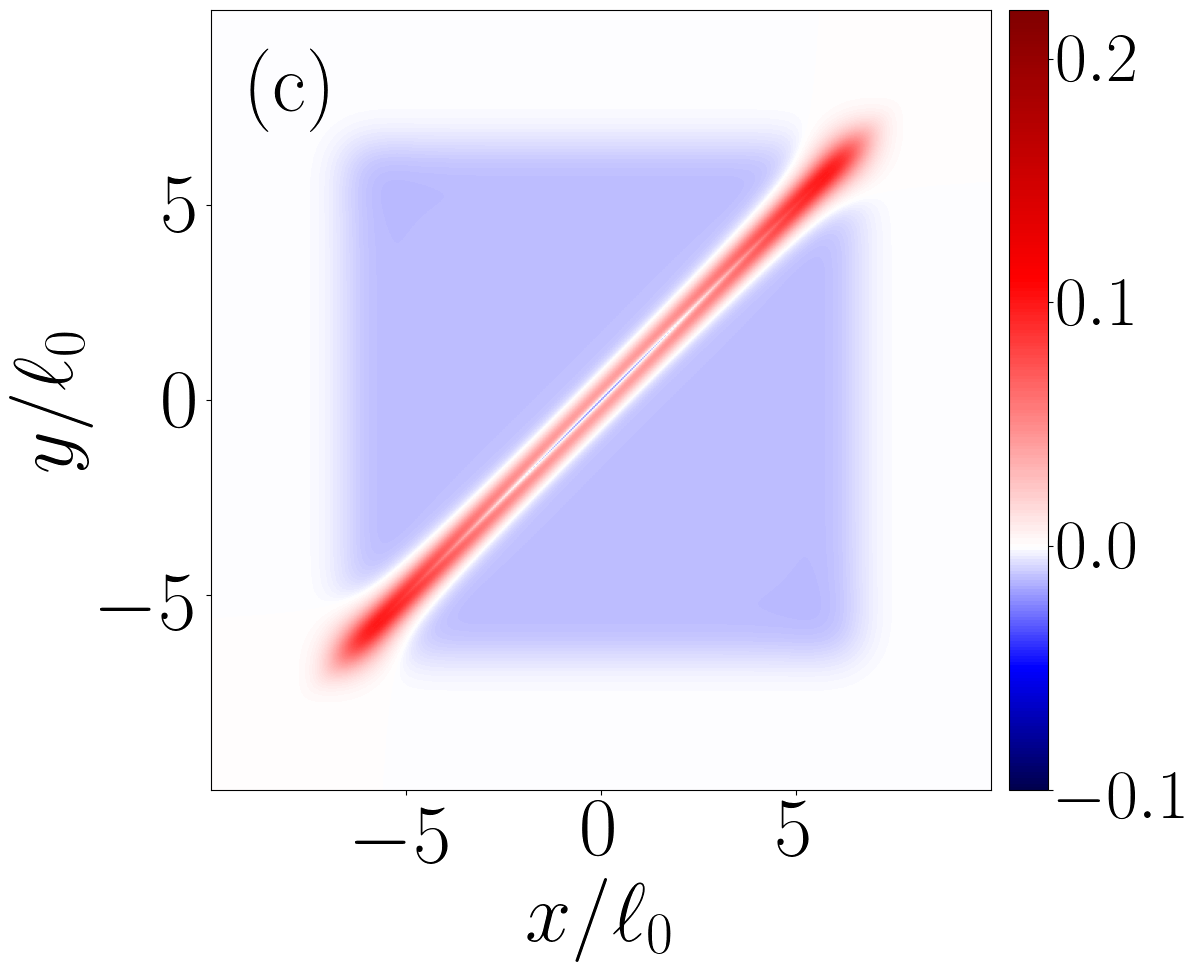}}}\qquad
{\subfigure
{\includegraphics[width=0.3\textwidth]{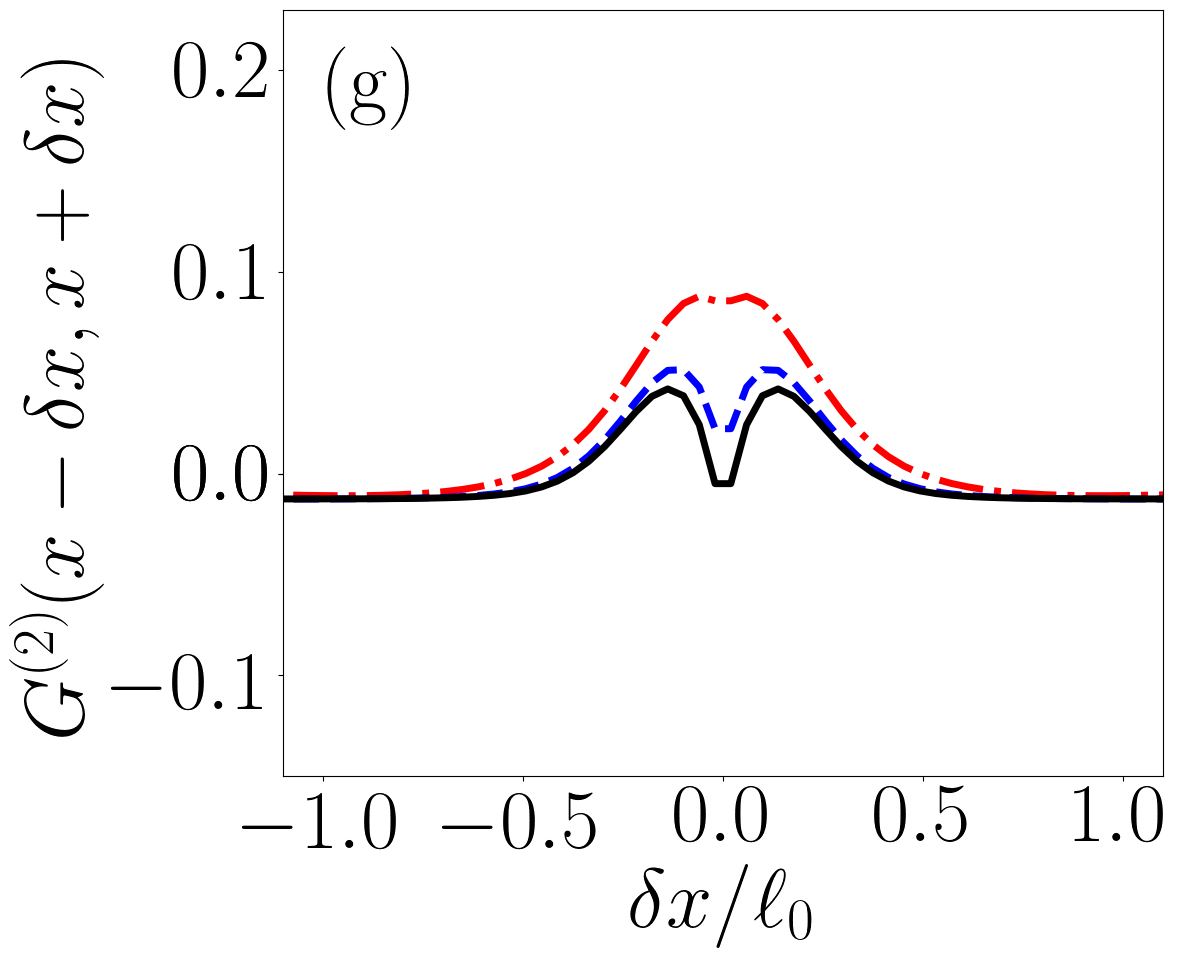}}}\\
{\subfigure
{\includegraphics[width=0.3\textwidth]{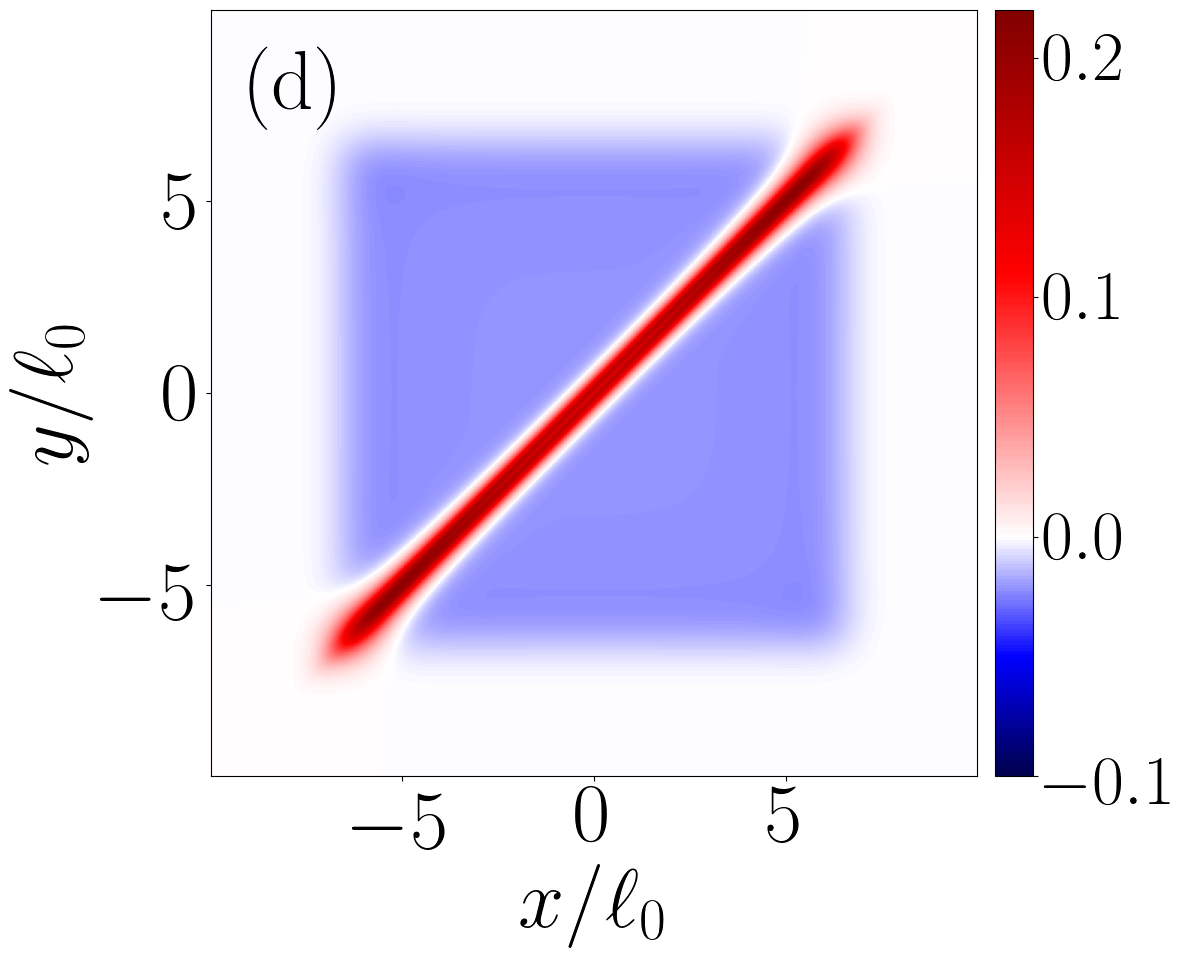}}}\qquad
{\subfigure
{\includegraphics[width=0.3\textwidth]{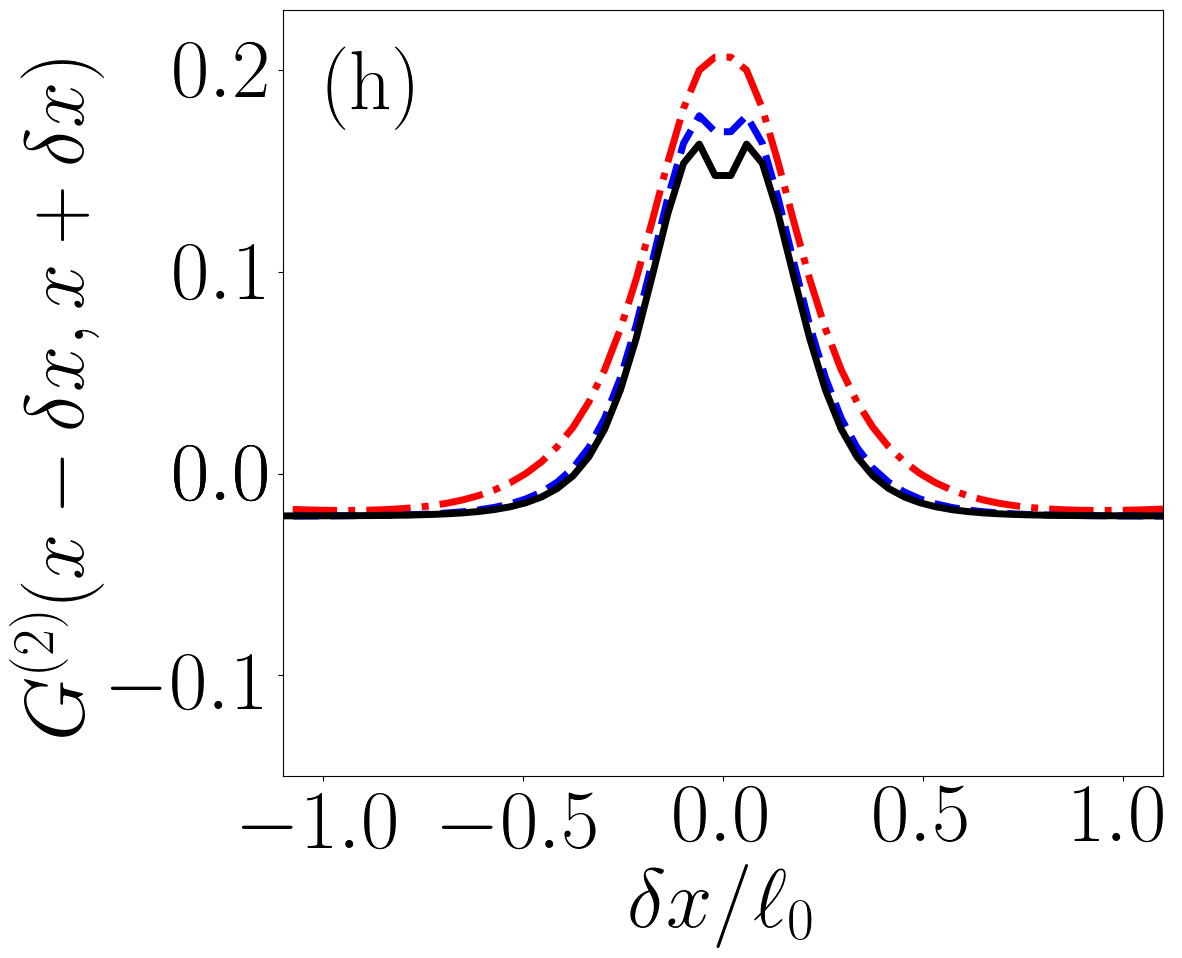}}}
\caption{Numerical solution for the two-body correlation $G_{\Phi\Lambda}^{(2)}(x_1,x_2)$ in real space. The panels (a-d) on the left column show colorplots of this quantity for growing values of the temperature, $K_B T/\mu=0,\,1,\,1.5,\,2.5$. The panels in the right column  show cuts $G_{\Phi\Lambda}^{(2)}(x-\delta x,x+\delta x)$ along straight lines parallel to the antidiagonal located at different spatial positions $x=0$ (black solid line) $x=2.5\ell_0$ (blue dashed line) $x=5\ell_0$ (red dot-dashed line) for the same values of the temperature.
Numerical calculations have been performed on the same grid as in Fig.\ref{Fig:1}.}
\label{Fig:2}
\end{figure*}

\section{Two-body correlations in a spatially homogeneous system\label{SubSec:Homogeneous}}

As a first application of this approach, in this Section we consider the simplest case of a homogeneous system of density $n_0$ and finite spatial size $L$ with periodic boundary conditions, for which an analytical solution to the Bogoliubov problem in Eq.~\eqref{Eq:BogOperator} is available. 

Thanks to the translation invariance of the system, solutions for $u_k(x)$ and $v_k(x)$ in a plane-wave form can be found with $u_k(x)=L^{-1/2}\,u_k\,\exp(ikx)$,  $v_k(x)=L^{-1/2}\,v_k\,\exp(ikx)$, and
\begin{equation}
	u_k=\sqrt{\frac{1}{2}\left(\frac{\xi_k}{\epsilon_k}+1\right)},\qquad
	v_k=-\sqrt{\frac{1}{2}\left(\frac{\xi_k}{\epsilon_k}-1\right)}.
	\label{Eq:uv}
\end{equation}
Here, we have used the short-hands $\xi_k=\epsilon_k^0+n_0 g_{1D}$, $\epsilon_k=\sqrt{\xi_k^2-(n_0 g_{1D})^2}$, and $\epsilon_k^0={\hbar^2 k^2}/{2m}$.

In position space, the order $\mathcal{O}(N)$ term of the correlation function is proportional to
\begin{multline}
G_{{\Phi_0}\Lambda}^{(2)}(x_1,x_2)=-\frac{1}{L}+\frac{1}{L}\sum_{k\neq 0}\left\{\left(e^{ik(x_1-x_2)}+e^{-ik(x_1-x_2)}\right)\right.\\
\left.\times\left[\left(u_k+v_k\right)^2(1+2N_k)-1\right]\right\}\, ,
\label{Eq:g1x}
\end{multline}
where we have exploited the condition  $N_k=N_{-k}$ imposed by the inversion symmetry of the system. Here, the first $-1/L$ term arises because of the fixed number of atoms $N$ that are in the system, and corresponds to the first term in Eq.~\eqref{Eq:Ggen_pL}.

In the momentum space, we focus instead on the $G_{\Lambda\Lambda}^{(2)}(k_1,k_2)$ term that can be decomposed in a normal and anomalous part according to \eqref{Eq:Wick}. The normalized normal correlation takes the form
\begin{multline}
	g^{(2)}_{\Lambda\Lambda,{\rm N}}(k_1,k_2)\equiv\frac{G^{(2)}_{\Lambda\Lambda,{\rm N}}(k_1,k_2)}{G^{(1)}(k_1,k_1)G^{(1)}(k_2,k_2)}=\\
	=1+\delta_{k_1k_2}\frac{\left|N_{k_1} u_{k_1}^2+\left(1+N_{k_1}\right)v_{k_1}^2\right|^2}{G^{(1)}(k_1,k_1)G^{(1)}(k_2,k_2)}=1+\delta_{k_1k_2}.
	\label{Eq:g2N}
\end{multline}
In a spatially homogeneous configuration, the populations of the different momentum components are in fact independent, so this correlation function differs from unity only for $k_1=k_2$. Its peak value on the diagonal is equal to 2, as typical for thermal states. This is also true for the case of the quantum depleted atoms at $T=0$, as first noticed in \cite{Cayla-expHBT-2020} and explained in the following. In the next Section, we will see how this bunching peak broadens in a trapped geometry, while keeping a maximum value equal to 2.

Analogously, the normalized anomalous correlation function writes
\begin{equation}
\begin{split}
	g^{(2)}_{\Lambda\Lambda,{\rm A}}(k_1,k_2)&\equiv\frac{G^{(2)}_{\Lambda\Lambda,{\rm A}}(k_1,k_2)}{G^{(1)}(k_1,k_1)G^{(1)}(k_2,k_2)}\\
		&=\frac{\left|(1+2N_{k_1})u_{k_1}v_{k_2}\right|^2}{G^{(1)}(k_1,k_1)G^{(1)}(k_2,k_2)}\,\delta_{k_1,-k_2}\, ,
\end{split}
\end{equation}
which is different from zero only for $k_1=-k_2$. This condition can be physically understood from the properties of the quantum depletion underlying the anomalous correlation at $T=0$: Since the particles belonging to the quantum depletion form a coherent state of pairs of particles at opposite momenta virtually ejected out of the condensate by the interactions, the anomalous correlations are non-zero only for $k_1=-k_2$. 

In contrast to the normal correlations, the peak value of $g^{(2)}_{\Lambda\Lambda,{\rm A}}(k_1,k_2)$ for $k_1=-k_2$ has a non-trivial $k$-dependence. In particular, since $u_k\to 1$ and $v_k\to 1/[2(\xi k)^2]\to 0$ in the $k\to\infty$ limit (according to Eqs.~\eqref{Eq:uv}), the peak correlation diverges for $k\to\infty$. In the simplest $T=0$ case, one has $N_k=0$ and thus $g^{(2)}_{\Lambda\Lambda,{\rm A}}(k,-k)\to 4(\xi k)^4$. In the more general $T>0$ case, the large-$k$ divergence has the same form, but it is visible only at high enough values of the momentum for which $N_k\lesssim 1$. Also, in this case, the finite spatial extension of the condensate results in a broadening of this correlation feature.

\section{Two-body correlations in a harmonically trapped system\label{Sec:Harmonic}}

The detailed review of the spatially homogeneous case presented in the previous Section paves the way to the study of the correlations in the harmonically trapped system that we are now going to explore. In the first Subsec.~\ref{SubSec:PosSpace} we investigate the position space two-body correlations. In the following Subsec.~\ref{SubSec:MomSpace}, we provide an in-depth study of the momentum-space correlations.

\begin{figure*}[!t]
\centering
{\subfigure
{\includegraphics[width=0.32\textwidth]{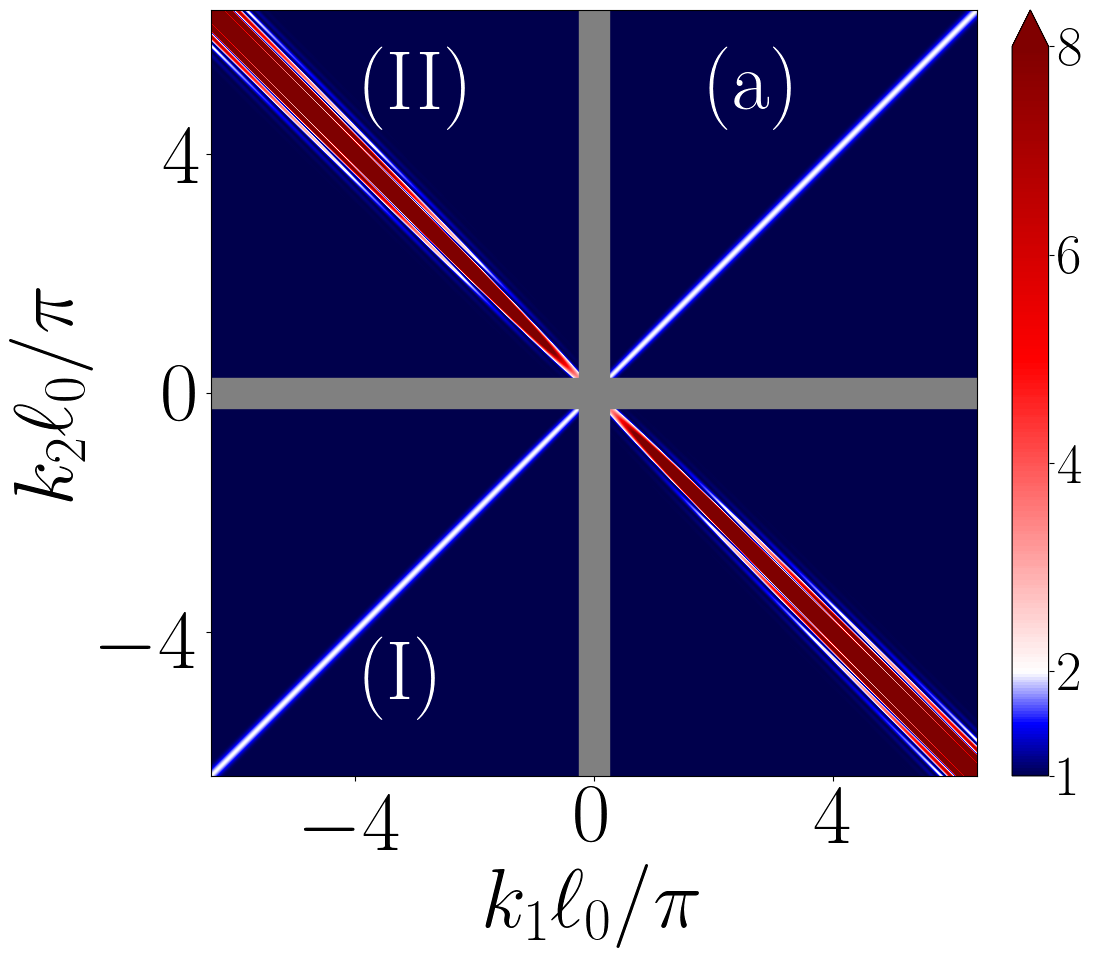}}}
{\subfigure
{\includegraphics[width=0.32\textwidth]{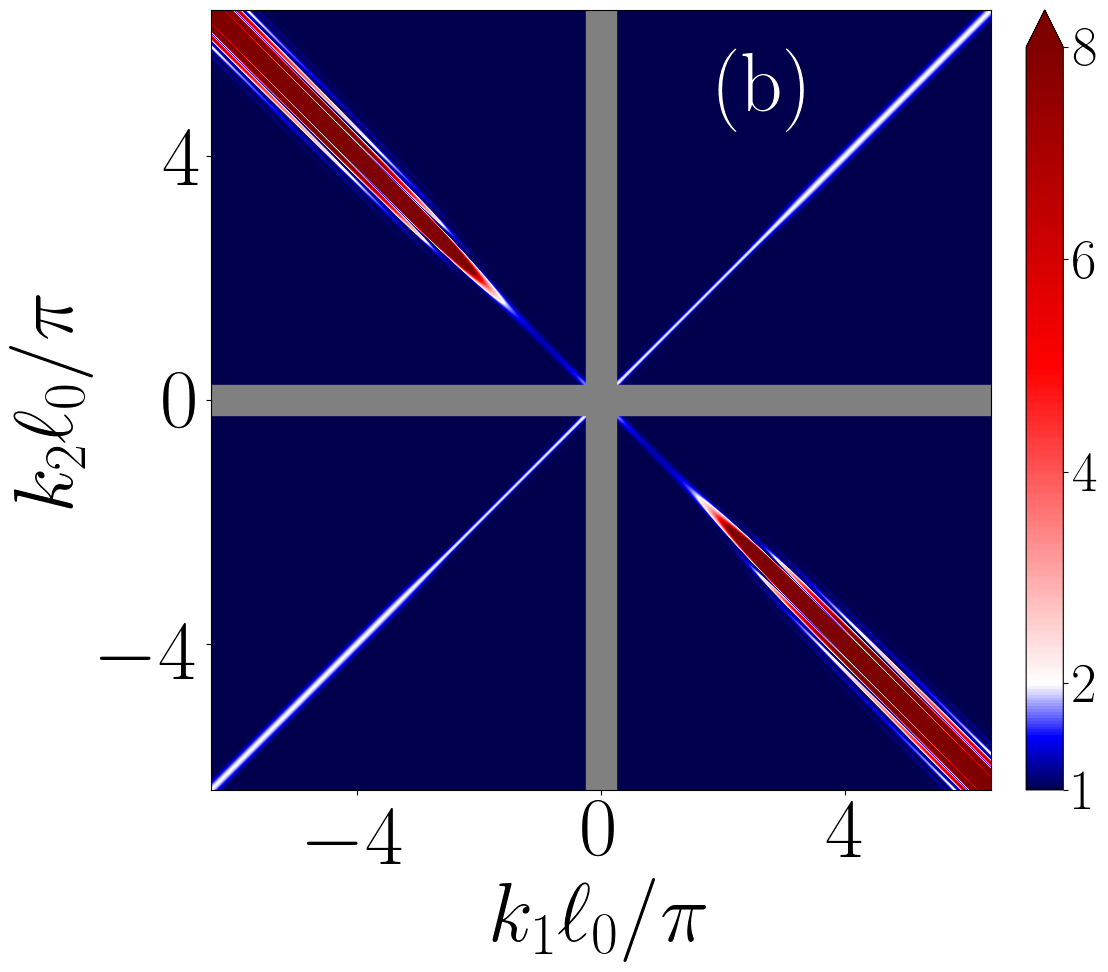}}}
{\subfigure
{\includegraphics[width=0.32\textwidth]{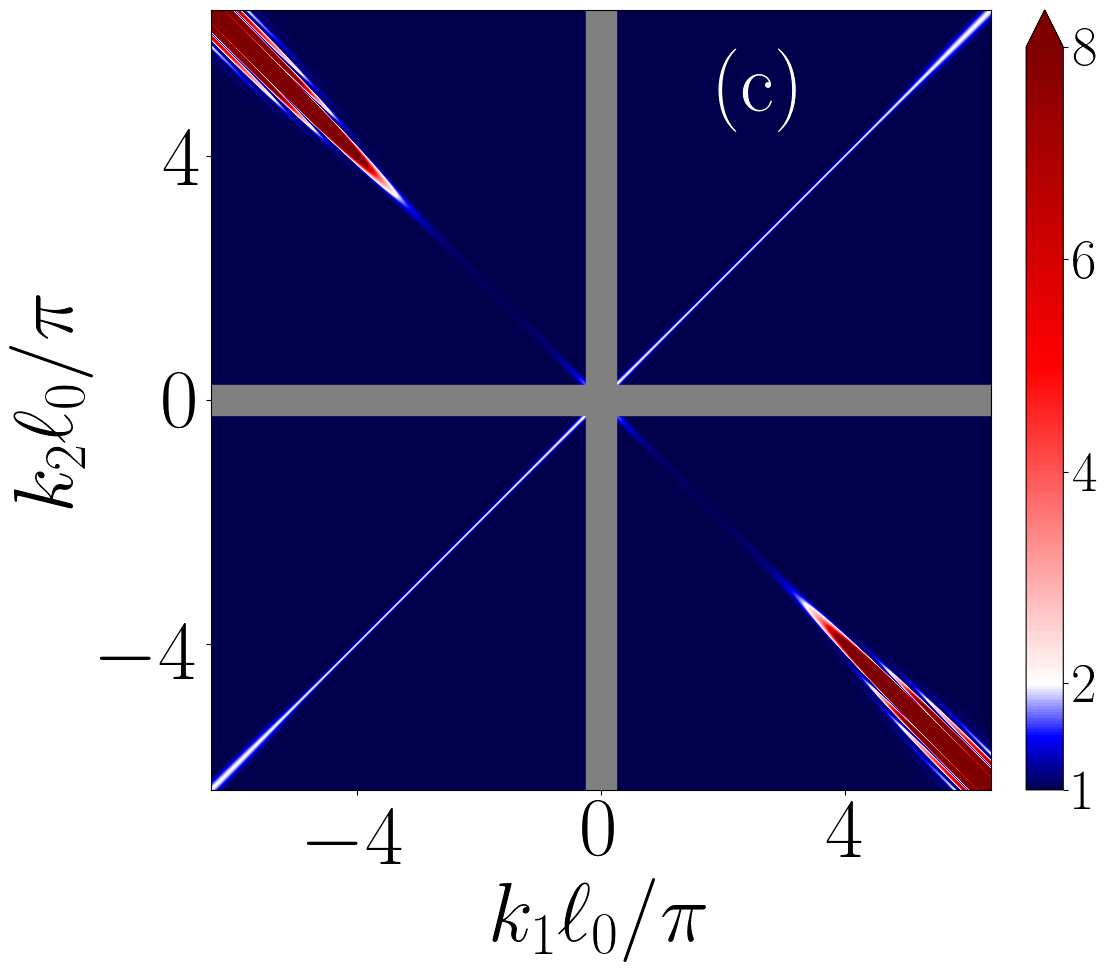}}}
\caption{ Numerical results of the two-body correlation function $g_{\Lambda\Lambda}^{(2)}(k_1,k_2)$ in momentum space. Panels (a-c) show the profiles of $g_{\Lambda\Lambda}^{(2)}(k_1,k_2)$ for the values of the temperatures $K_B T/\mu=0,\,1,\,2.5$. The white vertical and horizontal stripes, of width $k\ell_0=0.5\pi$, identify the region of the momentum space where the condensate contribution to the correlations is located. This region is not relevant for our purposes. We notice the presence of two main features in the two-body correlations in momentum space: (I) the diagonal stripe accounts for the normal contribute $g_{\Lambda\Lambda,{\rm N}}^{(2)}(k_1,k_2)$ to the fourth order correlator; (II) the anti-diagonal stripe, account instead for the anomalous contribute $g_{\Lambda\Lambda,{\rm A}}^{(2)}(k_1,k_2)$ to the fourth order correlator.}
\label{Fig:3}
\end{figure*}

\subsection{Position-space correlations\label{SubSec:PosSpace}}

The plots in Figs.~\ref{Fig:2}(a-d) illustrate the modification of the density-density correlations captured by the two-body correlation function $G^{(2)}_{\Phi_0\Lambda}(x_1,x_2)$ for increasing temperatures, $K_B T/\mu=0,\,1,\,1.5,\,2.5$. At $T=0$ a well-contrasted anti-bunching stripe is visible whose origin reflects the reduced probability of finding two atoms close by because of the repulsive interaction. The typical width of the anti-bunching stripe is given by the interaction strength and is of order the healing length. In the Bogoliubov theory, this feature is accounted for by zero-point vacuum contribution (i.e., for $N_k=0$) in Eq.~\eqref{Eq:g1x}. 

At finite temperature, in contrast, thermal fluctuations induce a bunching associated with their chaotic character. The width of this bunching bump, set by the temperature, is initially broader than the anti-bunching feature already present at $T=0$ and gets narrower for growing $T$. Upon increasing the temperature, the thermal fluctuations of the density therefore progressively overtake the (quantum) anti-bunching feature. When the temperature is much larger than the interaction energy $\mu$, the bunching ends up dominating over the anti-bunching dip which is no longer visible [Figs.~\ref{Fig:2}(g-h)].

The anti-bunching effect originating from the repulsive interaction and the bunching effect occurring at finite temperature change differently from one another along the spatial profile of the cloud. As it appears in the panels (e-h), the narrow anti-bunching dip has a larger amplitude at the center of the cloud (solid black lines) where the density is largest and the effect of repulsive interactions strongest. The inhomogeneous density profile also results in a smaller value of the healing length at the center, and a correspondingly narrower dip in the two-point correlations [Fig.~\ref{Fig:2}(e)]. On the other hand, at low temperatures [Fig.~\ref{Fig:2}(b,f)] the broad bunching bump starts being visible and has the strongest impact in the spatial region outside the condensate where the thermal atoms are mostly located, as shown in the previous Fig.\ref{Fig:1}(d). In the bottom panels Fig.\ref{Fig:2}(d,h) for the highest temperature $k_B T =2.5\mu$, a substantial thermal component is present also at the trap center, so the overall width of the bunching peak is approximately constant throughout the cloud.

\subsection{Momentum-space correlations\label{SubSec:MomSpace}}

The plots in Fig.~\ref{Fig:3}(a-c) show examples of the two-body correlations in the momentum-space at different temperatures. In typical experiments~\cite{Cayla-expHBT-2020}, these momentum space correlations are extracted as two-body correlations after a long time-of-flight expansion from the source of trapped atoms. In doing this, care must be paid that the signal is not distorted by interaction effects during the  time-of-flight expansion. 
Here, we calculate the normalized contribution that dominates once the momenta $k_1$ and $k_2$ are taken outside the condensate region (see Section~II),
\begin{equation}
    g_{\Lambda\Lambda}^{(2)}(k_1,k_2)=\frac{G_{\Lambda\Lambda}^{(2)}(k_1,k_2)}{G^{(1)}(k_1,k_1)\,G^{(1)}(k_2,k_2)}.
\end{equation} 
The condensate region is indicated by the vertical and horizontal grey stripes in Fig.~\ref{Fig:3}(a-c). 

In analogy with the analytical results discussed in Sec.~\ref{SubSec:Homogeneous} for the homogeneous case, two characteristic features can be identified in these plots: (I) a positive correlation along the $k_1\approx k_2$ diagonal,  due to the normal average of particle-particle correlations $g_{\Lambda\Lambda,{\rm N}}^{(2)}(k_1,k_2)$ ; (II) a positive correlation along the $k_1\approx -k_2$ anti-diagonal, due to the anomalous average of particle-particle correlations $g_{\Lambda\Lambda,{\rm A}}^{(2)}(k_1,k_2)$. In contrast to the homogeneous, infinite case, both these features are here broadened by the finite size of the system. In particular the $k$-dependence of the linewidth provides detailed information on the microscopic physics of the fluid.

\subsubsection{Normal averages\label{SubSubSec:NormalAv}}

The positive correlation signal along the $k_1\simeq k_2$ diagonal in Fig.~\ref{Fig:3}(a-c) can be interpreted as a HB-T phenomenon. It indeed corresponds to  the bunching of bosons with chaotic (thermal) statistics  and is  characterized by a remarkably uniform value of the peak amplitude $g^{(2)}(k_1,k_2=k_1)=2$ for all values of $k$.
This value is in quantitative agreement with the usual HB-T picture, but, as  recently pointed out in \cite{Cayla-expHBT-2020}, the underlying physics is more subtle than the usual HB-T of non-interacting bosons. 

The bunching phenomenon results in fact from two distinct contributions, that of the thermally excited Bogoliubov  modes (at $T\neq0$) and that of the quantum depletion in the Bogoliubov ground-state. The chaotic character of the statistics associated with those two contributions has a different physical origin. On the one hand, the Bogoliubov excitation modes are non-interacting bosons whose population has a thermal distribution, and their statistics is therefore the well-known chaotic (thermal) statistics of ideal bosons. On the other hand, for the quantum depletion, the chaotic character results from the destruction of its quantum coherence when the correlations are probed locally, $k_1\simeq k_2$, in the momentum-space. Indeed, while the quantum depletion is formed by pairs of particles at opposite momenta, the local two-body correlations probe the statistics of particles belonging to two different pairs, discarding the second partner of each particles in a pair.

\begin{figure}[!htbp]
\centering
{\subfigure
{\includegraphics[width=0.3\textwidth]{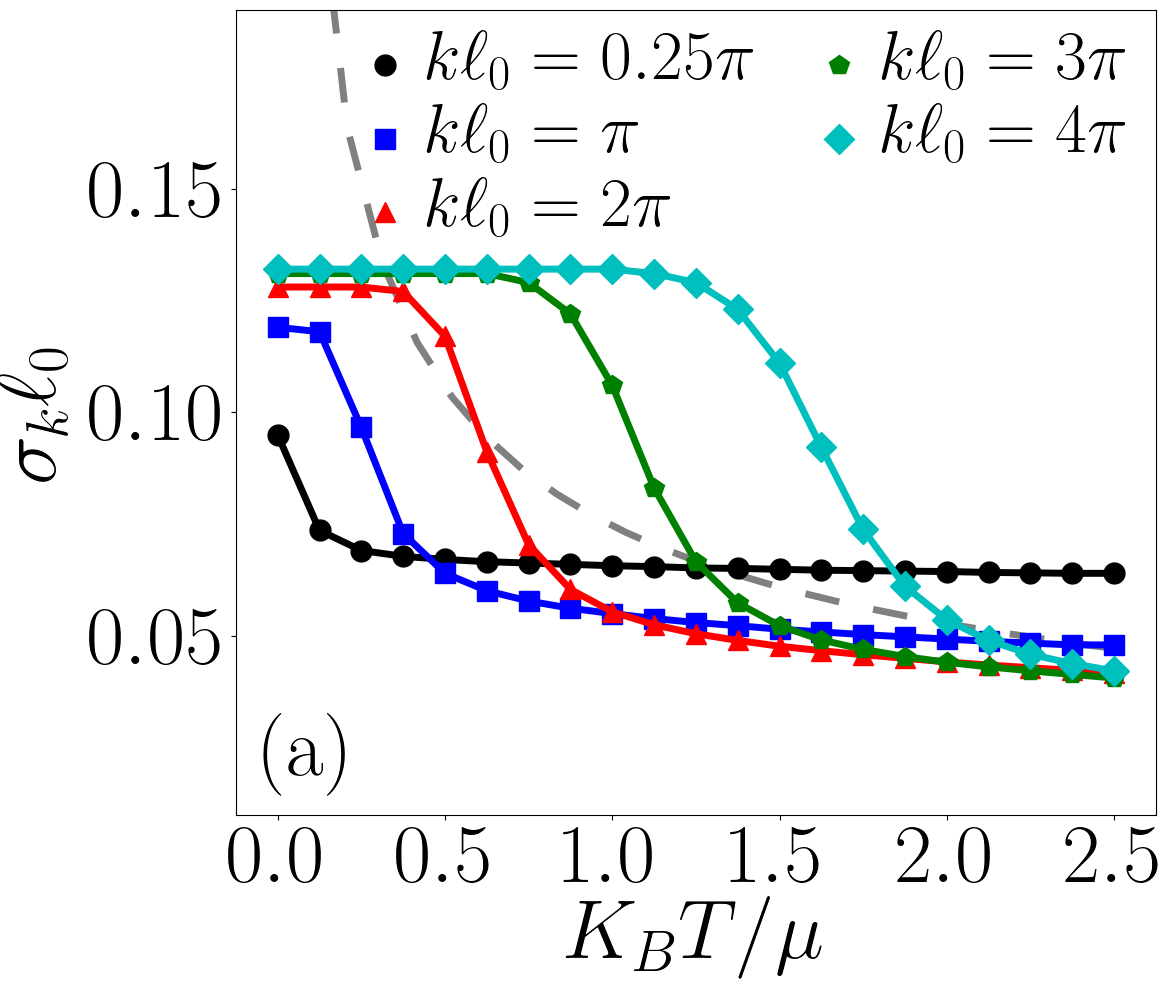}}}\\
{\subfigure
{\includegraphics[width=0.3\textwidth]{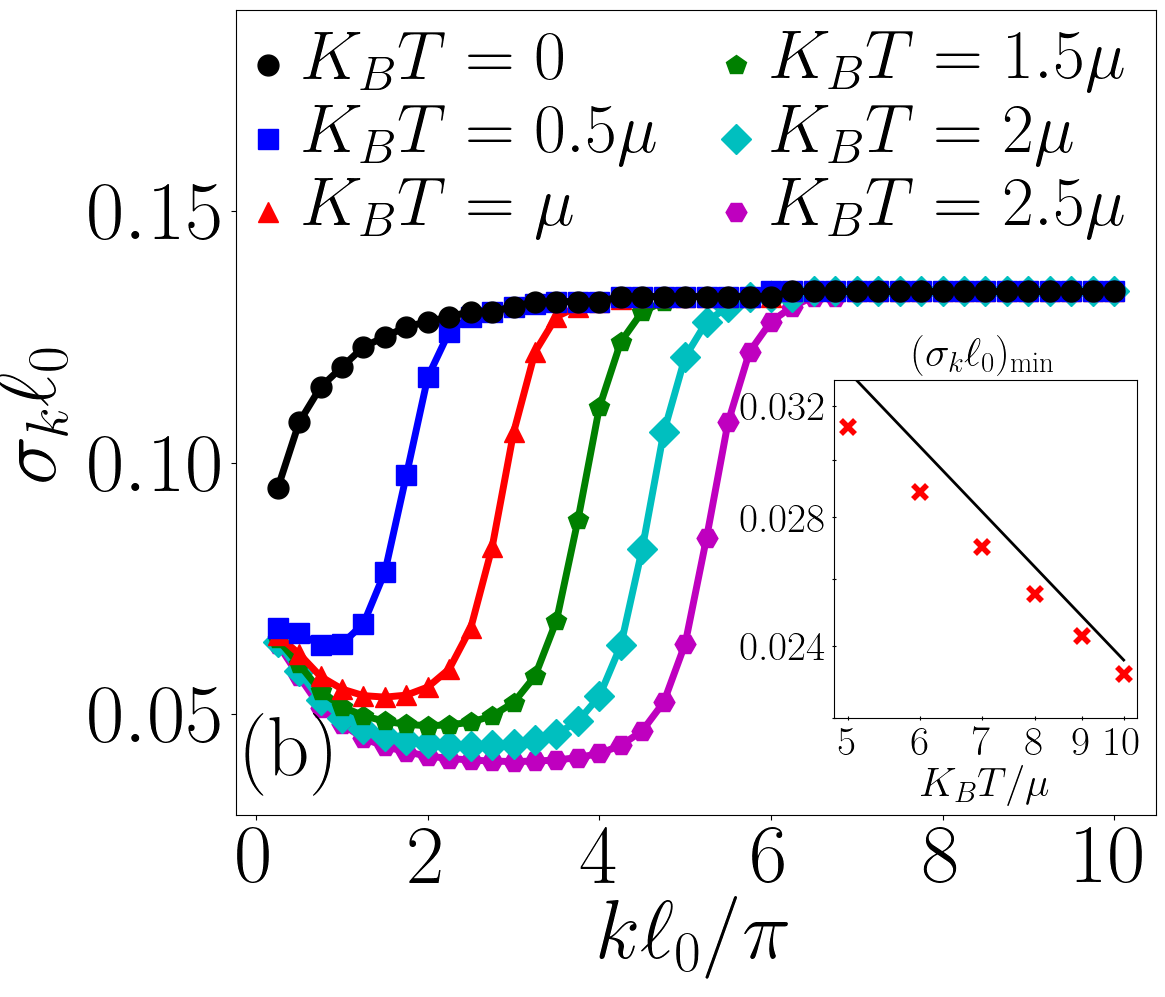}}}
\caption{Numerical results of the normalized two-body correlation function $g_{\Lambda\Lambda}^{(2)}(k_1,k_2)$ in momentum space. Panel (a) shows how the rms of the cuts $g_{\Lambda\Lambda,{\rm N}}^{(2)}(k+\delta k,k-\delta k)$ through the diagonal feature varies with the temperature, for the values of $k\ell_0/\pi=0.25,\,1,\,2,\,3,\,4$. At the higher temperatures (and for high enough momentum values so  that the Bose-Einstein statistics can be approximated by the Boltzmann one) the curves asymptotically follow the typical $\ell_0 \sqrt{2k_B T/ \hbar \omega_x}$ dependence of the non-degenerate, harmonically trapped, gas (gray-dashed line). Panel (b) shows how the rms of the cuts $g_{\Lambda\Lambda,{\rm N}}^{(2)}(k+\delta k,k+\delta k)$ varies with the wave vector $k$, for the values of the temperature $K_B T/\mu=0,\,0.5,\,1,\,1.5,\,2,\,2.5$. The inset shows the scaling with the temperature of the value of the plateau (red markers), compared with the analytical result in Eq.~\eqref{eq:inv_sigma_x} relative to the non-interacting, classical gas.}
\label{Fig:4}
\end{figure}

\begin{figure}[!htbp]
\centering
{\subfigure
{\includegraphics[width=0.27\textwidth]{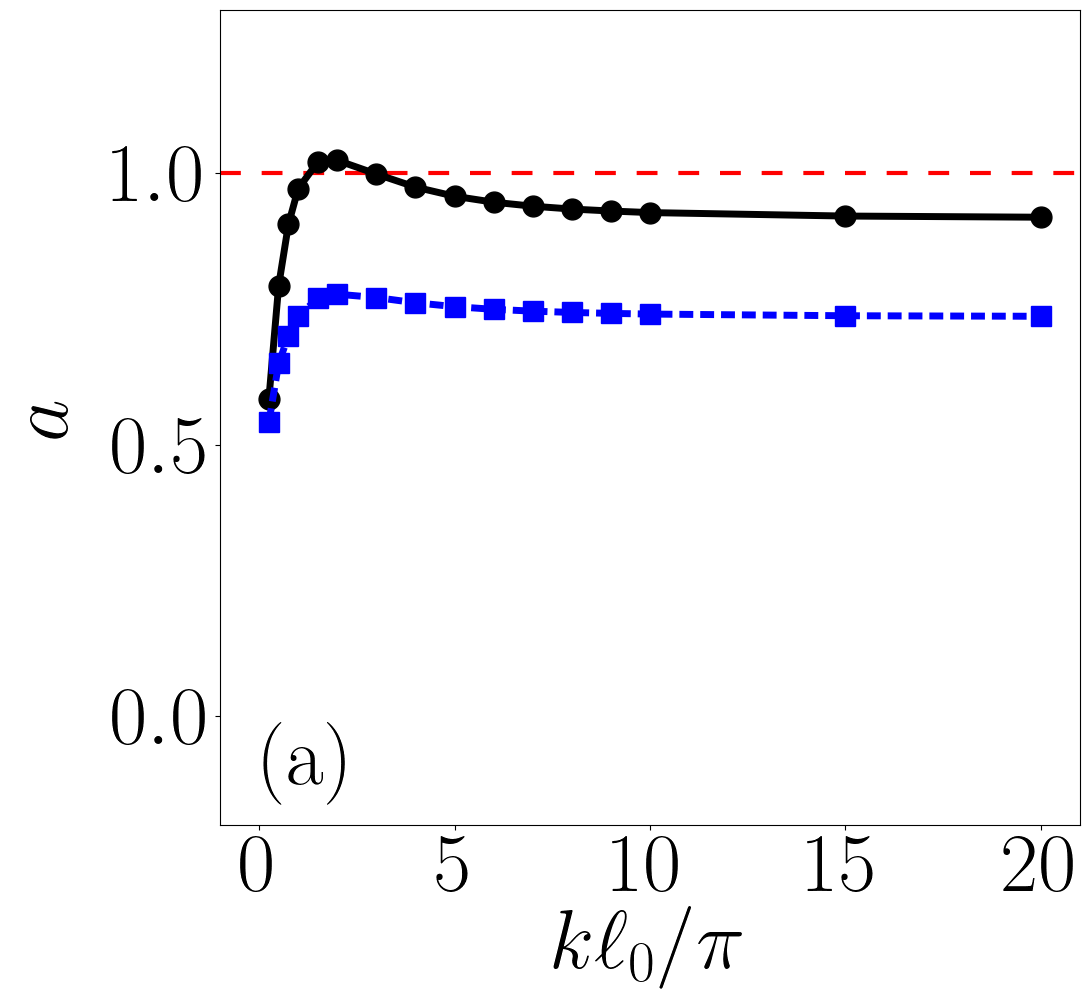}}}\\
{\subfigure
{\includegraphics[width=0.27\textwidth]{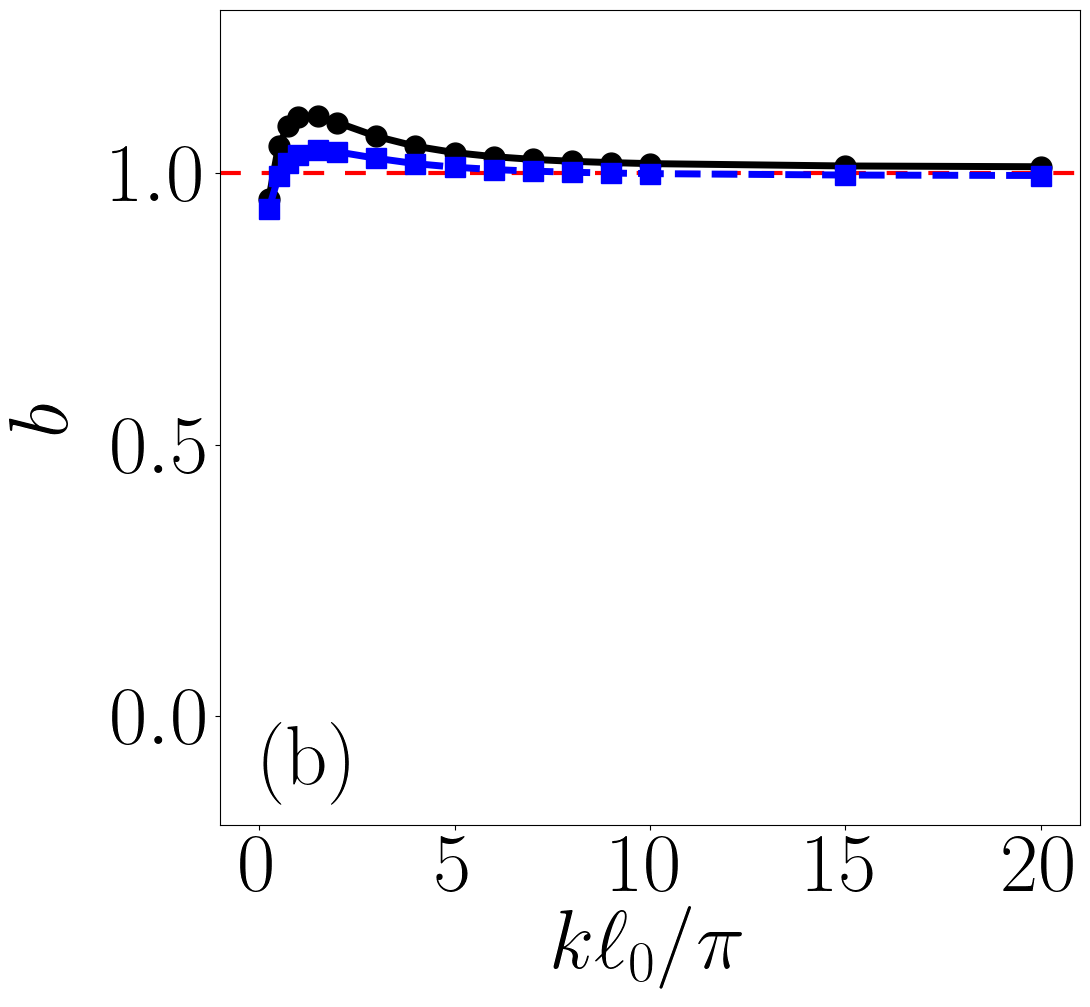}}}
\caption{Scaling of the rms of the cuts $g_{\Lambda\Lambda,{\rm N}}^{(2)}(k+\delta k,k-\delta k)$ through the diagonal (Black) and $g_{\Lambda\Lambda,{\rm A}}^{(2)}(k+\delta k,k+\delta k)$ through the anti-diagonal (blue) features with the size $L_{\rm bec}$ of the condensate at $T=0$. In panels (a) and (b) we respectively show the exponent $b$ and the prefactor $a$ of the fit of the numerical data with the power law $a/L_{\rm bec}^b$.}
\label{Fig:6}
\end{figure}

\begin{figure}[!bhtp]
\centering
{\subfigure
{\includegraphics[width=0.3\textwidth]{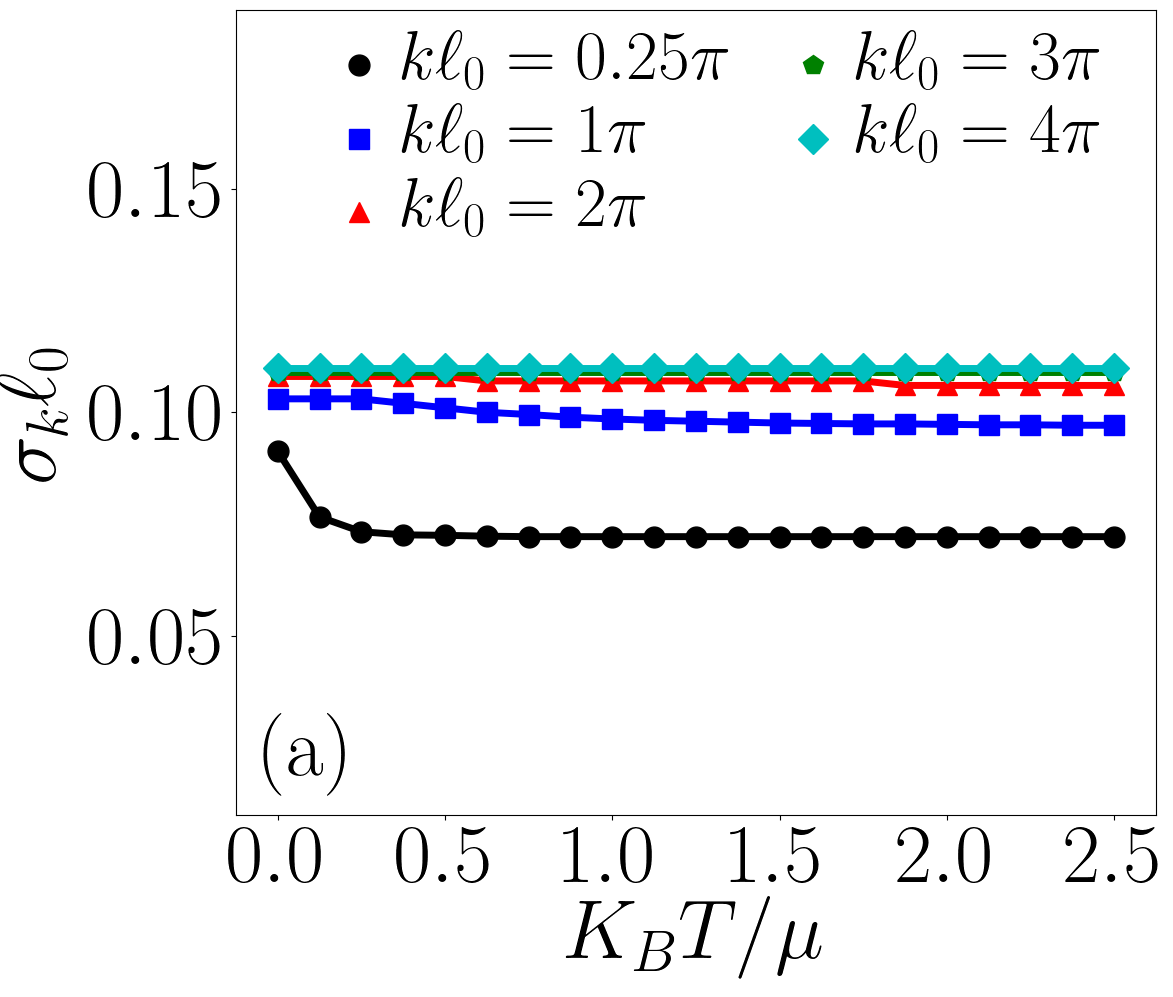}}}\\
{\subfigure
{\includegraphics[width=0.3\textwidth]{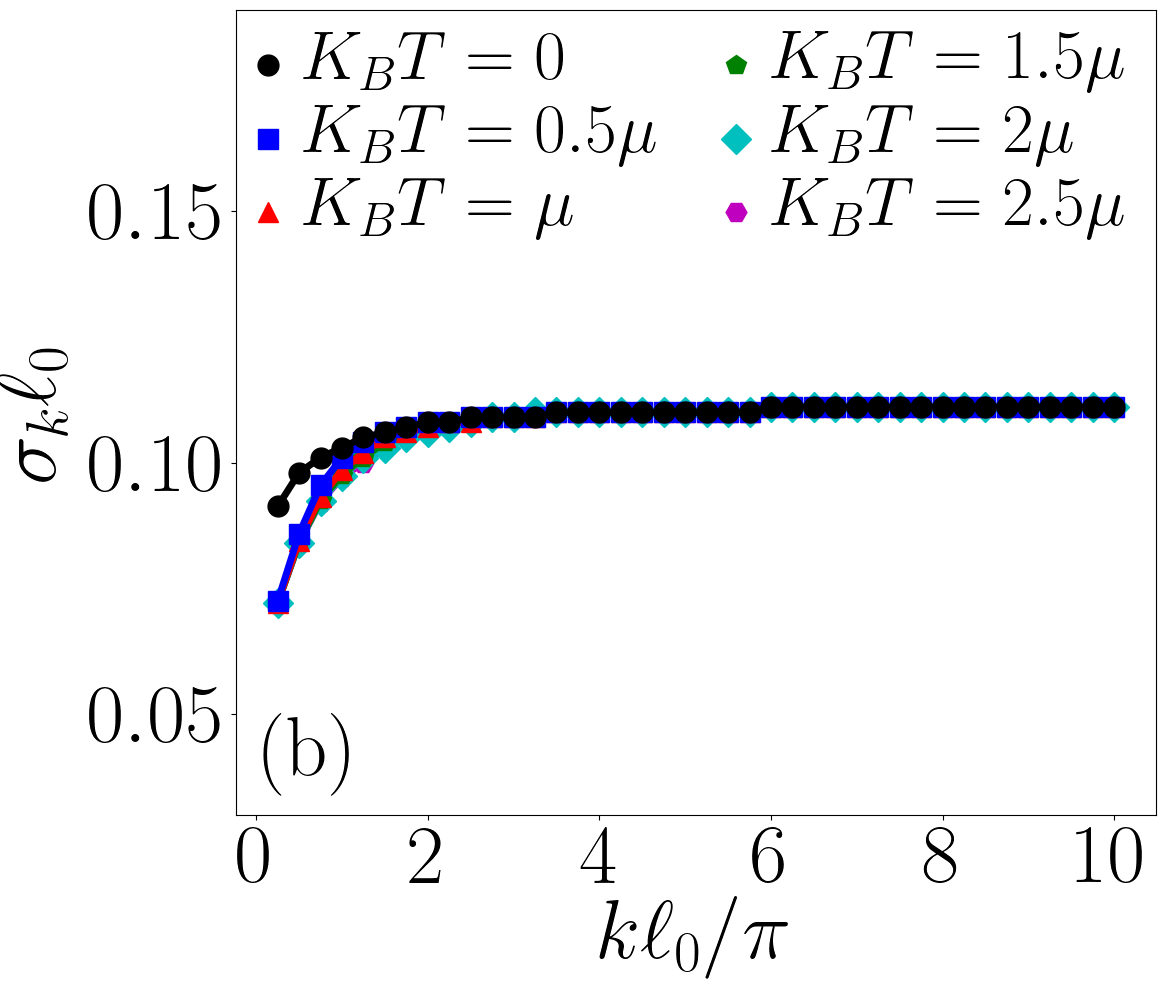}}}
\caption{Numerical results of the normalized two-body correlation function $g_{\Lambda\Lambda}^{(2)}(k_1,k_2)$ in momentum space. Panels (a,b) show the rms of the cuts $g_{\Lambda\Lambda,{\rm N}}^{(2)}(k+\delta k,k+\delta k)$ through the anti-diagonal feature for the same values of the parameters as in Fig.~\ref{Fig:4}.}
\label{Fig:5}
\end{figure}

Even richer is the dependence of the linewidth $\sigma_k$ of this HB-T peak on the system temperature $T$ and on the specific position in $k$ at which the linewidth is measured. 
Extending the usual HB-T argument to our more complex configuration, we can anticipate that the width of the bunching peak is inversely proportional to the spatial size of the components that provide the strongest contribution at position $k$ at temperature $T$. With this general trend in mind, we now turn to the detailed discussion of the numerical results shown in Figs.~\ref{Fig:4}-\ref{Fig:5}. More specifically, we plot there the root-mean-square width $\sigma_k$ of the HB-T peak. For any values $k$ and $T$, $\sigma_k$ is extracted from the analysis of cuts along the line $k_1+k_2=2 k$ in the two-dimensional plots of $g^{(2)}(k_1,k_2)$ at temperature $T$, like those in Fig,~\ref{Fig:3}(a-c).

\paragraph{Temperature-dependence}
At zero or low temperatures, the width $\sigma_k$ is dominated by the contribution from the quantum depletion of the condensate. For the relatively large value $\mu/\hbar\omega_x=11.2$ considered in the figures, the quantum depletion spreads over many Bogoliubov modes of the trapped system.
The spatial size of the quantum depletion is determined by the Bogoliubov functions $v_n(x)$ in Eq.~\eqref{eq:N(1)}, which are non-zero only in the region of the condensate. As a consequence, the width $\sigma_k$ tends to a finite value in the $T\to 0$ limit, inversely proportional to the condensate size $L_{\rm bec}$ (defined as the Thomas-Fermi radius). For the parameters of  Fig.~\ref{Fig:4}(a), this corresponds to the value $\sigma_k \ell_0 \approx 0.13$. Quite remarkably, this corresponds to a value of $\sigma_k L_{\rm bec}\approx 0.9$ which hints at a $1/L_{\rm bec}$ dependence. This feature will be further investigated in Fig.~\ref{Fig:6}. 

On the other hand, at high temperatures the thermal component described by the Bogoliubov functions $u_n(x)$ in Eq.~\eqref{eq:N(1)} dominates. The spatial size of these functions increases with the mode index $n$ and ends up extending well beyond the condensate size. This provides the decrease of the width $\sigma_k$ with the temperature $T$ that is visible in Fig.~\ref{Fig:4}(a). More quantitatively, the overall rms size of the thermal cloud of a harmonically trapped non-interacting non-degenerate gas follows a
\begin{equation}
    \sigma_x=\sqrt{\frac{k_B T} {m\omega_x^2}}=\ell_0 \sqrt{\frac{2k_B T}{\hbar \omega_x}}
\end{equation} 
dependence. Taking into account the explicit form of the one-body density matrix of the harmonically trapped, non-degenerate gas~\cite{landau5}
\begin{multline}
\rho^{(1)}(x_1,x_2)=
\exp\left[-\frac{m\omega_x^2}{8k_B T} (x_1+x_2)^2\right]\\ \exp\left[-\frac{m k_B T}{2\hbar^2} (x_1-x_2)^2\right],
\end{multline}
where we have assumed that $k_B T \gg \hbar \omega_x$, and transforming this expression to momentum space
\begin{multline}
    \rho^{(1)}(k_1,k_2)=\exp\left[-\frac{\hbar^2}{8mk_B T} (k_1+k_2)^2\right] \\ \exp\left[-\frac{k_B T}{2m\omega_x^2} (k_1-k_2)^2\right],
\end{multline}
we obtain the explicit expression
\begin{equation}
    \sigma_k=\sqrt{\frac{m\omega_x^2}{8k_B T}}=\frac{1}{\sqrt{8}\,\sigma_x}
    \label{eq:inv_sigma_x}
\end{equation}
for the momentum space linewidth~\footnote{Note that the momentum space linewidth is defined as the rms linewidth for $\Delta k$ along the $(k-\Delta k,k+\Delta k)$ curve in the $(k_1,k_2)$ plane.}, which confirms the inverse proportionality on the spatial size of the system expected from the HB-T picture. This curve is displayed here as a dashed line and is found to accurately capture the numerical results for high enough temperatures. The corrections that are visible on the small $k$ curves are due to the quantum degeneracy of the thermal occupation of the lowest modes. 

The two low- and high-temperature regimes are separated by a sharp transition. The position of the transition is determined by the value of the temperature for which the thermal population of the Bogoliubov excitation mode starts dominating over the quantum depletion. As expected, the transition point moves towards higher values of $T$ for growing $k$ since the energy of the Bogoliubov eigenmodes giving the dominant contribution increases with $k$.

\paragraph{Wavevector-dependence} 
In Fig.~\ref{Fig:4}(b), we plot the width $\sigma_k$ as a function of the momentum $k$ where it is evaluated, for different temperatures. Our interpretation associating the HB-T width to the spatial size of the corresponding $k$-component is confirmed. At zero temperature $T=0$ (black), $\sigma_k$ slightly increases with the momentum $k$, which can be interpreted in terms of the stronger localization of the $v_n(x)$ functions of the high-$n$ Bogoliubov modes contributing to the quantum depletion, as shown in Fig.\ref{Fig:1}(c).

At non-zero temperatures, the curves exhibit a non-monotonic and richer behavior with $k$. At large $k$, all the curves at different temperatures collapse on the $T=0$ result for $\sigma_k$. This happens because the high $k$ components are not thermally populated as their energy is larger than the temperature, and the only contribution is that from the quantum depletion. 

As $k$ decreases one observes a sharp decrease of $\sigma_k$ followed by a plateau. This sharp jump corresponds to the sharp transition between the low- and high-temperature regimes identified above in the $T$-dependent analysis. Indeed, for momenta $k$ whose Bogoliubov energy is sufficiently small for the thermal population to dominate over the quantum depletion, the dominant contribution comes from the $u_n(x)$ functions which extend beyond the condensate and provide a wide thermal cloud of (almost) non-interacting particles as discussed above. This interpretation is confirmed by the close to $1/\sqrt{T}$ scaling of the value of $\sigma_k$ on the plateau displayed in the inset of Fig.\ref{Fig:4}(b).

For even smaller values of $k$, the plateau ends and $\sigma_k$ displays a slight increase for decreasing $k$. This feature can be connected to  the phononic character of such modes as well as to the corrections to the $\sigma_k\propto 1/\sqrt{T}$ dependence due to the quantum degeneracy of the low-lying non-condensed modes, as already noticed in Fig.\ref{Fig:4}(a).

\paragraph{System-size-dependence} 
To further reinforce our conclusions that the HB-T width $\sigma_k$ is inversely proportional to the spatial size of the corresponding $k$-component, we now explicitly study the dependence of the width $\sigma_k$ with the condensate size $L_{\rm bec}$. 
In the high-$T$ case, analytical insight on the inverse proportionality on the size of the thermal cloud was provided in Eq.\eqref{eq:inv_sigma_x}. Here, we focus on the bunching of the quantum depletion in the interacting $T=0$ case, where the condensate size $L_{\rm bec}$ is controlled by varying the strength of interactions, that is the chemical potential. 

For different values of the cut position $k$, we fit the dependence of $\sigma_k$ on $L_{\rm bec}$ with a power law of the form $a/L_{\rm bec}^b$, with the results plotted in Fig.~\ref{Fig:6} (black line). Except for a relatively small deviation for small values of $k$, the result in the panel \ref{Fig:6}(b) confirms the expected $L^{-1}_{\rm bec}$ dependence for values of $k\ell_0\gtrapprox 5\pi$. Comparison with Fig.~\ref{Fig:4}(b) shows that this value is comparable to that at which the width $\sigma_k$ reaches the asymptotic value discussed in the previous section. 

\subsubsection{Anomalous averages\label{SubSubSec:AnomalousAv}}

Fig.\ref{Fig:5} shows an analogous analysis for the positive correlation bump located on the anti-diagonal $k_1+k_2=0$ and associated with the anomalous averages. According to Eq.~\eqref{eq:A(1)}, this feature is determined by the products of $u_n(x)$ and $v_n(x)$ Bogoliubov amplitudes and the overall spatial size is determined by the latter. As we have seen in Fig.\ref{Fig:1}(c), the $v_n(x)$ roughly follow the condensate shape for all Bogoliubov modes, so we do not expect a marked temperature dependence for the width $\sigma_k$. 
This physical picture is confirmed by the numerical results shown in Fig.~\ref{Fig:5} (a) for different positions $k$ of the cut, which clearly display a much weaker $T$-dependence as compared to the corresponding curves in Fig.~\ref{Fig:4}(a). 

The same physics is also visible in the $k$-dependent curves shown in Fig.~\ref{Fig:5}(b), which feature an analogously weak dependence for all considered temperatures. 
In contrast to the complex behaviours observed for the diagonal feature studied in Fig.~\ref{Fig:4}(b), the $k$-dependence is here  a monotonically growing one and can be again interpreted in terms of the weaker localization of the $v_n(x)$ functions for the lower modes displayed in Fig.\ref{Fig:1}.

An explicit illustration of the condensate-size dependence at $T=0$ is provided by the blue curves in Fig.\ref{Fig:6}. Also in this case, for sufficiently large values of $k$, the scaling in panel (b) points to an inverse proportionality of the momentum-space linewidth on the condensate size, $\sigma_k \propto 1/L_{\rm bec}$. The smaller value of the prefactor $a$ visible in panel (a) can be interpreted in terms of the dependence of the anti-diagonal features on $v_n(x)$ rather than the $v_n(x)^2$ one of the quantum depletion.

\section{Conclusion\label{Sec:End}}

In this work, we have taken inspiration from the recent experiment in~\cite{Cayla-expHBT-2020} to carry out a detailed study of the consequences of the spatial inhomogeneity of a harmonically trapped Bose-Einstein condensate on the two-body correlation functions in both position and momentum spaces at low temperatures. To reduce the technical challenge and be able to unravel the basic physics of the system, we have made use of the Bogoliubov description of the weakly interacting gas and we have restricted our attention to the computationally easier one-dimensional geometry. 
We are of course aware that the severe assumptions made in our work prevent us from quantitatively capturing all features of the experiment. Nonetheless, our calculations for an idealized model provide a crucial step in view of unraveling the subtle interplay of the finite spatial size with interaction and temperature effects.  

As a first step, we have unveiled intriguing features in the spatial structure of the Bogoliubov excitation modes, which result from a subtle interplay of the inhomogeneous density profile and the collective vs. single-particle character of the mode in the different spatial regions. 
This understanding of the excitation modes provides a crucial tool to investigate the two-body correlations in both the position and the momentum spaces.

In the position space, we find markedly different behaviours in the center of the trap and in the outer region: in the central region, the zero-temperature anti-bunching due to interactions in the condensate is slowly replaced by bunching of thermal atoms as temperature grows. The outer regions are instead dominated by thermal atoms that show thermal bunching at any finite temperatures.

In the momentum space, two main features are clearly visible in the correlation pattern. On the main diagonal, namely for $k_1 \approx k_2$, a marked Hanbury Brown and Twiss bunching signal is visible with the normalized correlation function going up to the usual value $g^{(2)}=2$ of chaotic fields, as experimentally observed in~\cite{Cayla-expHBT-2020}. Depending on the specific values of the momentum and  the temperature under consideration, the origin of this chaotic character is however very different. For thermally occupied modes, it is related to the Bogoliubov description of the thermal cloud in terms of non-interacting quasi-particles. For modes that are only populated by the quantum depletion, the thermal character originates from tracing out the opposite momentum states with which each state is  quantum correlated. 
This physical difference results in a different scaling of the linewidth of the HB-T bunching feature in different regions of $k$-space at a given temperature: for low-$k$, thermally occupied modes, this is determined by the inverse spatial size of the thermal cloud; for the high-$k$ modes occupied by the quantum depletion only, the linewidth is determined by the inverse  condensate size.

Another bunching feature appears on the antidiagonal for $k_1\approx -k_2$. This feature is due to the anomalous average of the atomic field operator, that is the quantum correlation between pairs of opposite momentum atoms forming the quantum depletion. In contrast to the diagonal feature, the antidiagonal one gets monotonically weaker for increasing temperatures without changing its qualitative shape. In all regimes, its linewidth is in fact fixed by the size of the condensate and displays  a weak temperature dependence.

Our natural next steps will consist in looking for the $k_1\approx -k_2$ features and exploiting the $k$-dependence to get deeper insight on the structure of the atomic gas in upgraded experiments along the lines of~\cite{Cayla-expHBT-2020}. Future theoretical work includes the investigation of correlations in regimes of higher temperatures and/or stronger interactions where excitations can no longer be considered as non-interacting quasi-particles and non-Gaussian corrections to the Bogoliubov theory must be included. All these studies will eventually contribute to establishing Hanbury Brown and Twiss techniques as a powerful experimental window on the microscopic physics of strongly correlated quantum gases.

\section{Acknowledgements}
S. B. acknowledges funding from the Leverhulme Trust Grant No.~ECF-2019-461 and the Lord Kelvin/Adam Smith (LKAS) Leadership Fellowship. I.C. acknowledges funding from Provincia Autonoma di Trento and from the Quantum Flagship Grant PhoQuS (820392) of the European Union. D. C. acknowledges support from the LabEx PALM (Grant number ANR-10-LABX-0039), the R\'egion Ile-de-France in the framework of the DIM SIRTEQ, the ``Fondation d'entreprise iXcore pour la Recherche" and the Agence Nationale pour la Recherche (Grant number ANR-17-CE30-0020-01).

\bibliography{HBT.bib}
\bibliographystyle{unsrt}

\end{document}